\documentclass[aps,physrev,reprint]{revtex4-2}
\usepackage{amsmath}
\usepackage{bm}
\usepackage{braket}
\usepackage{graphicx}
\usepackage{hyperref}
\usepackage{siunitx}
\renewcommand{\Re}{\operatorname{Re}}
\renewcommand{\Im}{\operatorname{Im}}
\DeclareMathOperator{\tr}{tr}
\DeclareMathOperator{\diag}{diag}
\providecommand{\abs}[1]{\lvert#1\rvert}
\providecommand{\norm}[1]{\lVert#1\rVert}
\begin{document}
\title{Small mode volume topological photonic states in one-dimensional lattices with dipole--quadrupole interactions}
\author{Raymond P. H. Wu}
\author{H.C. Ong}
\email[]{hcong@phy.cuhk.edu.hk}
\affiliation{Department of Physics, The Chinese University of Hong Kong, Sha Tin, New Territories, Hong Kong}
\date{\today}
\begin{abstract}
    We study the topological photonic states in one-dimensional (1-D) lattices analogue to the Su-Schrieffer-Heeger (SSH) model beyond the dipole approximation.
    The electromagnetic resonances of the lattices supported by near-field interactions between the plasmonic nanoparticles are studied analytically with coupled dipole--quadrupole method.
    The topological phase transition in the bipartite lattices is determined by the change of Zak phase.
    Our results reveal the contribution of quadrupole moments to the near-field interactions and the band topology.
    It is found that the topological edge states in non-trivial lattices have both dipolar and quadrupolar nature.
    The quadrupolar edge states are not only orthogonal to the dipolar edge states, but also spatially localized at different sublattices.
    Furthermore, the quadrupolar topological edge states, which coexist at the same energy with the quadrupolar flat band have shorter localization length and hence smaller mode volume than the conventional dipolar edge states.
    The findings deepen our understanding in topological systems that involve higher-order multipoles, or in analogy to the wave functions in quantum systems with higher-orbital angular momentum, and may be useful in designing topological systems for confining light robustly and enhancing light-matter interactions.
\end{abstract}
\maketitle
\section{\label{seci}Introduction}
Topological insulators are class of matter which are insulating in the bulk but conducting at the boundary with the backscattering-immune states that are robust against local perturbations\cite{hasan2010}.
The concepts of topological phases are not restricted to fermionic systems, but they also can be realized in bosonic and classical waves systems\cite{wang2019}.
In particular, topological photonics\cite{ozawa2019} complements the electronic counterpart and has been theoretically proposed\cite{haldane2008,wang2008,raghu2008} and experimentally realized\cite{wang2009} in two-dimensional (2-D) photonic crystals.
The Su-Schrieffer-Heeger (SSH) model\cite{su1979}, which originates from the study of soliton in polyacetylene, is the simplest system demonstrating non-trivial topological bands.
The photonic analogue of SSH model has been realized in photonic crystals\cite{keil2013,xiao2014}, chains of plasmonic\cite{poddubny2014,ling2015,sinev2015,downing2017,zhang2018,pocock2018,downing2018} and dielectric\cite{slobozhanyuk2015,slobozhanyuk2016,kruk2017} nanoparticles, and gyromagnetic lattices\cite{rphwu2019}.

In 1-D systems, the topological edge states existing within the band gaps are localized at the boundary of the system with distinct topological phases\cite{asboth2016}.
The spatial confinement of light by topological edge states enhanced light-matter interactions in subwavelength scale which leads to applications such as lasing\cite{st2017,zhao2018,parto2018} and sensing\cite{guo2021}.
Ideally, photonic states with high quality factor and small mode volume are desirable for such applications\cite{ryu2003,kippenberg2004,xiao2010,de2012,seidler2013,yang2015,wang2018,xwu2019}.
Recently, bound states in the continuum (BICs) in photonic systems are of great interest due to their infinitely high quality factor\cite{zhen2014,hsu2016,doeleman2018,pankin2020,azzam2021}.
In particular, the topological nature of BICs has been revealed\cite{zhen2014} and observed experimentally\cite{doeleman2018}.
However, topological BICs are different from the topological edge states in the way that while the former originates from the topological charges in the polarization vectors of the far-field radiation, the latter is from the closing of the band gaps arising from the mismatch between the band topologies of two physically joint bulk bands.
Currently, cavities made by plasmonic resonators are still the state of the art to obtain small mode volume\cite{kuttge2010,huang2016,hugall2018,epstein2020}.
On the other hand, the exponentially localized topological edge states in 1-D lattices may provide an alternative way to confine light in small mode volume while at the same time topologically protected.

Conventionally, the SSH model with dipole approximation is sufficient in studying the dipolar topological edge states in non-trivial systems\cite{poddubny2014,ling2015,slobozhanyuk2015,sinev2015,slobozhanyuk2016,kruk2017,downing2017,zhang2018,pocock2018,downing2018,rphwu2019}.
The solutions of the edge states under the dipole approximation have characteristic that the dipole moments are localized in only one of the sublattice sites.
This result is verified in several works\cite{ling2015,downing2017,zhang2018,pocock2018,downing2018,rphwu2019} including those where long-range interactions are included\cite{zhang2018,pocock2018,rphwu2019}.
The fields from the dipole moments are similar to the $sp^2$ hybridized orbital electron wave functions in the polyacetylene.
Although dipolar topological edge states have been widely studied, there is a lack of studies on the topological states that involve higher-order multipoles, or in analogy to the wave functions in quantum systems with higher-orbital angular momentum.
As such quantum systems are hard to be realized, photonic crystals or metamaterials may provide a feasible platform for us to explore them.

Previously, the quadrupole dispersion in three-dimensional (3-D) lattices of plasmonic sphere is shown to be intrinsically anisotropic, which defies a simple isotropic effective medium description without spatial dispersion\cite{han2009}.
The coupling strength between quadrupole resonance and external electromagnetic waves can be on the same order of magnitude as the magnetic dipole\cite{han2009}.
In particular, it is shown that the quadrupolar resonance leads to large bandwidth in 1-D periodic arrays of plasmonic nanoparticles due to strong coupling\cite{alu2009}.
Recently, the multipolar resonances in 2-D lattices have been studied\cite{evlyukhin2012,swiecicki2017,babicheva2018,babicheva2019}.
The coupling between the dipolar modes and the quadrupolar modes gives rise to interesting physics such as lattice anapole effect\cite{babicheva2019}.
Furthermore, sensing applications is proposed due to the higher sensitivity of the diffractive quadrupole resonance than the dipole resonance\cite{evlyukhin2012}.

In this work, we study the 1-D plasmonic lattices analogue to the SSH model that go beyond the dipole approximation by including dipole--quadrupole interactions.
The electromagnetic resonances of the lattices by near-field interactions between the plasmonic nanoparticles are studied analytically with coupled dipole--quadrupole method.
Our results reveal the contribution of quadrupole moments in the near fields.
The topological phase transition in the bipartite lattices is demonstrated by calculating the Zak phase.
It is found that, the topological edge states in non-trivial lattices have both dipolar and quadrupolar nature.
Surprisingly, the quadrupole edge states are not only orthogonal to the dipole edge states, but also spatially localized at different sublattice.
Furthermore, the quadrupolar topological edge states, which coexist at the same energy with the quadrupolar flat band have shorter localization length and hence smaller mode volume than the conventional dipolar edge states.
Our findings may be useful in designing topological systems for confining light robustly and enhancing light-matter interactions.

This article is organized as follows.
In Sec.~\ref{secii}, the coupled-dipole-quadrupole method for a collection of nanoparticles is formulated.
In Sec.~\ref{seciii}, the geometry and material of the nanoparticles is discussed.
The analytical solutions of 1-D monopartite lattices are presented in Sec.~\ref{seciv}.
Then the topological phase transition in the bipartite lattices is demonstrated in Sec.~\ref{secv}.
Finally, in Sec.~\ref{secvi}, the topological edge states in non-trivial lattices are studied.

\section{\label{secii}Coupled dipole--quadrupole method}
\begin{figure}
    \includegraphics[width=8.6cm]{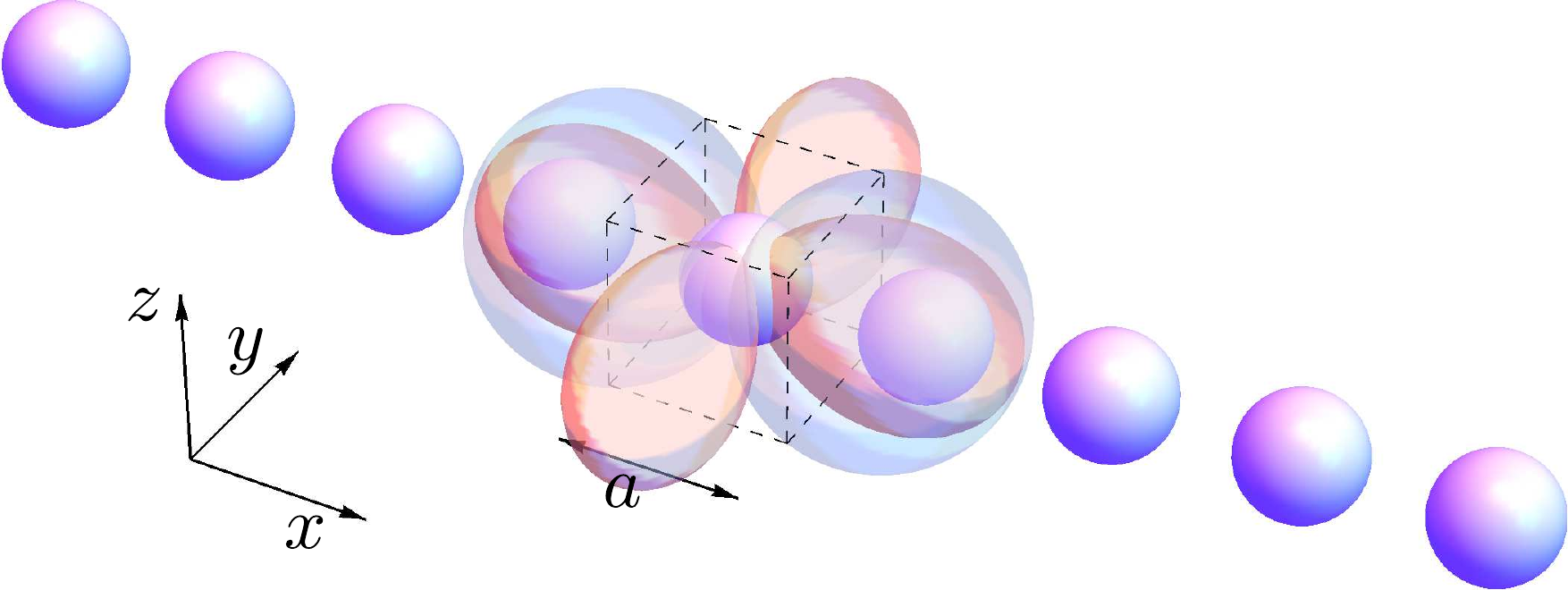}
    \caption{\label{fig1}Illustration of a collection of plasmonic nanoparticles on a 1-D lattice.
    The unit cell is indicated by the dashed box.
    Each sphere is approximated as a point dipole moment and a point quadrupole moment.
    The electric equipotentials of the dipole moment $p_x$ and the quadrupole moment $Q_{xx}$ for an nanoparticle are illustrated as blue and red surfaces, respectively, which correspond to the longitudinal modes.
    }
\end{figure}
We formulate the coupled dipole--quadrupole method by considering a collection of nanoparticles in air as depicted in Fig.~\ref{fig1}.
In the following, we work in Cartesian coordinates and the harmonic time dependence $e^{-i\omega t}$ is assumed and omitted.
Also, SI units are used throughout this article.
We approximate each nanoparticle at position $\bm{r}$ as a point electric dipole moment $\bm{p}(\bm{r})=(p_x,p_y,p_z)^T$ and a point electric quadrupole moment
\begin{equation}
    \bm{Q}(\bm{r})=
    \begin{pmatrix}
        Q_{xx}&Q_{xy}&Q_{xz}\\
        Q_{yx}&Q_{yy}&Q_{yz}\\
        Q_{zx}&Q_{zy}&Q_{zz}
    \end{pmatrix}\label{eqii1}
\end{equation}
at the center of the nanoparticle.
The quadrupole moment $\bm{Q}$ is traceless $\tr(\bm{Q})=0$ and symmetric $\bm{Q}=\bm{Q}^T$, such that only $5$ components, $Q_{xx}$, $Q_{xy}$, $Q_{xz}$, $Q_{yy}$, and $Q_{yz}$ are independent.
The induced dipole moment at $\bm{r}$ is given by
\begin{equation}
    \bm{p}(\bm{r})=\bm{\alpha}^p(\omega)\bm{E}(\bm{r}),\label{eqii2}
\end{equation}
and the induced quadrupole moment at $\bm{r}$ is given by\cite{han2009,alu2009}
\begin{equation}
    \bm{Q}(\bm{r})=\bm{\alpha}^Q(\omega)\left(\frac{\nabla\bm{E}(\bm{r})+\bm{E}(\bm{r})\nabla}{2}\right),\label{eqii3}
\end{equation}
where $\bm{\alpha}^p(\omega)$ is the dipole polarizability and $\bm{\alpha}^Q(\omega)$ is the quadrupole polarizability of the nanoparticles and we have
\begin{equation}
    (\nabla\bm{E}+\bm{E}\nabla)_{ij}=\frac{\partial E_j}{\partial i}+\frac{\partial E_i}{\partial j}.\label{eqii4}
\end{equation}
The electric field at $\bm{r}$ from a point dipole source at $\bm{r'}$ is given by\cite{novotny2012}
\begin{equation}
    \bm{E}^p(\bm{r})=\frac{k_0^2}{\varepsilon_0}\bm{G}^p(\bm{r},\bm{r'})\bm{p}(\bm{r'}),\label{eqii5}
\end{equation}
where $k_0=\omega/c$ is the wave number in the background medium and $\varepsilon_0$ is the permittivity.
The 3-D Green's tensor for a point dipole is given by\cite{evlyukhin2012,babicheva2018,babicheva2019}
\begin{widetext}
    \begin{equation}
        \bm{G}^p(\bm{r},\bm{r'})=\frac{e^{ik_0R}}{4\pi R}\left[\left(-\frac{1}{(k_0R)^2}+i\frac{1}{k_0R}+1\right)\bm{I}+\left(\frac{3}{(k_0R)^2}-i\frac{3}{k_0R}-1\right)\bm{\hat{n}}\otimes\bm{\hat{n}}\right],\label{eqii6}
    \end{equation}
\end{widetext}
where $\bm{I}$ is the second-order identity tensor and we define $\bm{R}(\bm{r},\bm{r'}):=\bm{r}-\bm{r'}$ and $\hat{\bm{n}}(\bm{r},\bm{r'}):=(\bm{r}-\bm{r'})/\abs{\bm{r}-\bm{r'}}$.
The components of $\bm{G}^p$ can be represented by $G^p_{ij}$, where $i$ and $j$ are Cartesian coordinates $x$, $y$, and $z$.
$\bm{G}^p$ is symmetric such that $\bm{G}^p=(\bm{G}^p)^T$.
On the other hand, the electric field at $\bm{r}$ from a point quadrupole source at $\bm{r'}$ is given by
\begin{equation}
    \bm{E}^Q(\bm{r})=\frac{k_0^2}{\varepsilon_0}\bm{G}^Q(\bm{r},\bm{r'})\bm{Q}(\bm{r'})\bm{\hat{n}}(\bm{r},\bm{r'}),\label{eqii7}
\end{equation}
and the 3-D Green's tensor for a point quadrupole is given by\cite{evlyukhin2012,babicheva2018,babicheva2019}
\begin{widetext}
    \begin{equation}
        \bm{G}^Q(\bm{r},\bm{r'})=\frac{e^{ik_0R}}{4\pi R^2}\left[\left(-\frac{1}{(k_0R)^2}+i\frac{1}{k_0R}+\frac{1}{2}-i\frac{k_0R}{6}\right)\bm{I}+\left(\frac{5}{2(k_0R)^2}-i\frac{5}{2k_0R}-1+i\frac{k_0R}{6}\right)\bm{\hat{n}}\otimes\bm{\hat{n}}\right].\label{eqii8}
    \end{equation}
\end{widetext}
Again, the components of $\bm{G}^Q$ can be represented by $G^Q_{ij}$ and $\bm{G}^Q$ is symmetric leading to $\bm{G}^Q=(\bm{G}^Q)^T$.
The superposition of $\bm{E}^p$, $\bm{E}^Q$, and the external excitation field $\bm{E}^0$ yields the total electric field at $\bm{r}$
\begin{equation}
    \bm{E}(\bm{r})=\bm{E}^0(\bm{r})+\bm{E}^p(\bm{r})+\bm{E}^Q(\bm{r}).\label{eqii9}
\end{equation}
Then the coupled dipole--quadrupole equations are given by
\begin{widetext}
    \begin{subequations}
        \label{eqii10}
        \begin{eqnarray}
            \bm{p}(\bm{r})&=&\bm{\alpha}^p\left(\bm{E}^0(\bm{r})+\sum_{\bm{r'}\neq\bm{r}}\frac{k_0^2}{\varepsilon_0}\bm{G}^p(\bm{r},\bm{r'})\bm{p}(\bm{r'})+\sum_{\bm{r'}\neq\bm{r}}\frac{k_0^2}{\varepsilon_0}\bm{G}^Q(\bm{r},\bm{r'})\bm{Q}(\bm{r'})\bm{\hat{n}}(\bm{r},\bm{r'})\right),\label{eqii10a}
            \\
            \bm{Q}(\bm{r})&=&\frac{\bm{\alpha}^Q}{2}\left(\nabla\bm{E}^0(\bm{r})+\bm{E}^0(\bm{r})\nabla\right)\nonumber
            \\*
            &&+\frac{\bm{\alpha}^Q}{2}\frac{k_0^2}{\varepsilon_0}\sum_{\bm{r'}\neq\bm{r}}\left[\nabla\left(\bm{G}^p(\bm{r},\bm{r'})\bm{p}(\bm{r'})\right)+\left(\bm{G}^p(\bm{r},\bm{r'})\bm{p}(\bm{r'})\right)\nabla\right]\nonumber
            \\*
            &&+\frac{\bm{\alpha}^Q}{2}\frac{k_0^2}{\varepsilon_0}\sum_{\bm{r'}\neq\bm{r}}\left[\nabla\left(\bm{G}^Q(\bm{r},\bm{r'})\bm{Q}(\bm{r'})\bm{\hat{n}}(\bm{r},\bm{r'})\right)+\left(\bm{G}^Q(\bm{r},\bm{r'})\bm{Q}(\bm{r'})\bm{\hat{n}}(\bm{r},\bm{r'})\right)\nabla\right].\label{eqii10b}
        \end{eqnarray}
    \end{subequations}
\end{widetext}

By expanding and rearranging terms, we transform Eq.~(\ref{eqii10}) to a system of linear equations\cite{evlyukhin2012}.
For a collection of $N\geq1$ nanoparticles with positions at $\bm{r}_n$ with $n=0,1,\ldots,N-1$, we define the state vector for each nanoparticle as
\begin{equation}
    \bm{X}_n=(p_x,p_y,p_z,Q_{xx},Q_{xy},Q_{xz},Q_{yy},Q_{yz})^T.\label{eqii11}
\end{equation}
Finally, we have
\begin{equation}
    (\bm{A}^{-1}(\omega)-\bm{\Gamma}(\omega))\bm{X}=\bm{F},\label{eqii12}
\end{equation}
where
\begin{equation}
    \bm{X}=
    \begin{pmatrix}
        \bm{X}_0\\
        \bm{X}_1\\
        \vdots\\
        \bm{X}_{N-1}\\
    \end{pmatrix},\label{eqii13}
\end{equation}
is the state vector, $\bm{A}(\omega)$ is the polarizability matrix, $\bm{\Gamma}(\omega)$ is the interaction matrix, and $\bm{F}$ is the external excitation field vector.

\subsection{\label{seciia}Infinite periodic lattices}
For infinite periodic lattices with position vector $\bm{R}$, the translational symmetry leads to
\begin{subequations}
    \label{eqiia1}
    \begin{eqnarray}
        \bm{p}(\bm{r}+\bm{R})&=&e^{i\bm{k}\cdot\bm{R}}\bm{p}(\bm{r}),\label{eqiia1a}
        \\
        \bm{Q}(\bm{r}+\bm{R})&=&e^{i\bm{k}\cdot\bm{R}}\bm{Q}(\bm{r}),\label{eqiia1b}
    \end{eqnarray}
\end{subequations}
where $\bm{k}$ is the Bloch wave vector.
Also, the dipolar and quadrupolar Green's tensors follow
\begin{subequations}
    \label{eqiia2}
    \begin{eqnarray}
        \bm{G}^p(\bm{r},\bm{r}+\bm{R})&=&\bm{G}^p(0,\bm{R}),\label{eqiia2a}
        \\
        \bm{G}^Q(\bm{r},\bm{r}+\bm{R})&=&\bm{G}^Q(0,\bm{R}).\label{eqiia2b}
    \end{eqnarray}
\end{subequations}
Then Eq.~(\ref{eqii10}) becomes
\begin{widetext}
    \begin{subequations}
        \label{eqiia3}
        \begin{eqnarray}
            \bm{p}(\bm{r})&=&\bm{\alpha}^p\left(\bm{E}^0(\bm{r})+\sum_{\bm{R}\neq0}\frac{k_0^2}{\varepsilon_0}\bm{G}^p(0,\bm{R})e^{i\bm{k}\cdot\bm{R}}\bm{p}(\bm{r})+\sum_{\bm{R}\neq0}\frac{k_0^2}{\varepsilon_0}\bm{G}^Q(0,\bm{R})e^{i\bm{k}\cdot\bm{R}}\bm{Q}(\bm{r})\bm{\hat{n}}(0,\bm{R})\right),\label{eqiia3a}
            \\
            \bm{Q}(\bm{r})&=&\frac{\bm{\alpha}^Q}{2}\left(\nabla\bm{E}^0(\bm{r})+\bm{E}^0(\bm{r})\nabla\right)\nonumber
            \\*
            &&+\frac{\bm{\alpha}^Q}{2}\frac{k_0^2}{\varepsilon_0}\sum_{\bm{R}\neq0}\left[\nabla\left(\bm{G}^p(0,\bm{R})e^{i\bm{k}\cdot\bm{R}}\bm{p}(\bm{r})\right)+\left(\bm{G}^p(0,\bm{R})e^{i\bm{k}\cdot\bm{R}}\bm{p}(\bm{r})\right)\nabla\right]\nonumber
            \\*
            &&+\frac{\bm{\alpha}^Q}{2}\frac{k_0^2}{\varepsilon_0}\sum_{\bm{R}\neq0}\left[\nabla\left(\bm{G}^Q(0,\bm{R})e^{i\bm{k}\cdot\bm{R}}\bm{Q}(\bm{r})\bm{\hat{n}}(0,\bm{R})\right)+\left(\bm{G}^Q(0,\bm{R})e^{i\bm{k}\cdot\bm{R}}\bm{Q}(\bm{r})\bm{\hat{n}}(0,\bm{R})\right)\nabla\right].\label{eqiia3b}
        \end{eqnarray}
    \end{subequations}
\end{widetext}
The $8N$ equations of Eq.~(\ref{eqii12}) is reduced to $8$, and we have
\begin{equation}
    (\bm{A}^{-1}(\omega)-\bm{\Gamma}(\bm{k},\omega))\bm{X}=\bm{F},\label{eqiia4}
\end{equation}

\subsection{\label{seciib}Quasi-electrostatic limit}
We will focus our study in the quasi-electrostatic limit where $k_0\to0$ such that Eq.~(\ref{eqii5}) and Eq.~(\ref{eqii7}), the electric fields at $\bm{r}$ from a point dipole source and a point quadrupole source at $\bm{r'}$, become
\begin{widetext}
    \begin{equation}
        \lim_{k_0\to0}\bm{E}^p(\bm{r})=\frac{1}{4\pi\varepsilon_0}\left[\left(-\frac{1}{R^3}\right)\bm{I}+\left(\frac{3}{R^3}\right)\bm{\hat{n}}\otimes\bm{\hat{n}}\right]\bm{p}(\bm{r'}),\label{eqiib1}
    \end{equation}
    and
    \begin{equation}
        \lim_{k_0\to0}\bm{E}^Q(\bm{r})=\frac{1}{4\pi\varepsilon_0}\left[\left(-\frac{1}{R^4}\right)\bm{I}+\left(\frac{5}{2R^4}\right)\bm{\hat{n}}\otimes\bm{\hat{n}}\right]\bm{Q}(\bm{r'})\bm{\hat{n}}(\bm{r},\bm{r'}).\label{eqiib2}
    \end{equation}
\end{widetext}

\section{\label{seciii}Geometry and material}
\begin{figure}
    \includegraphics[width=8.6cm]{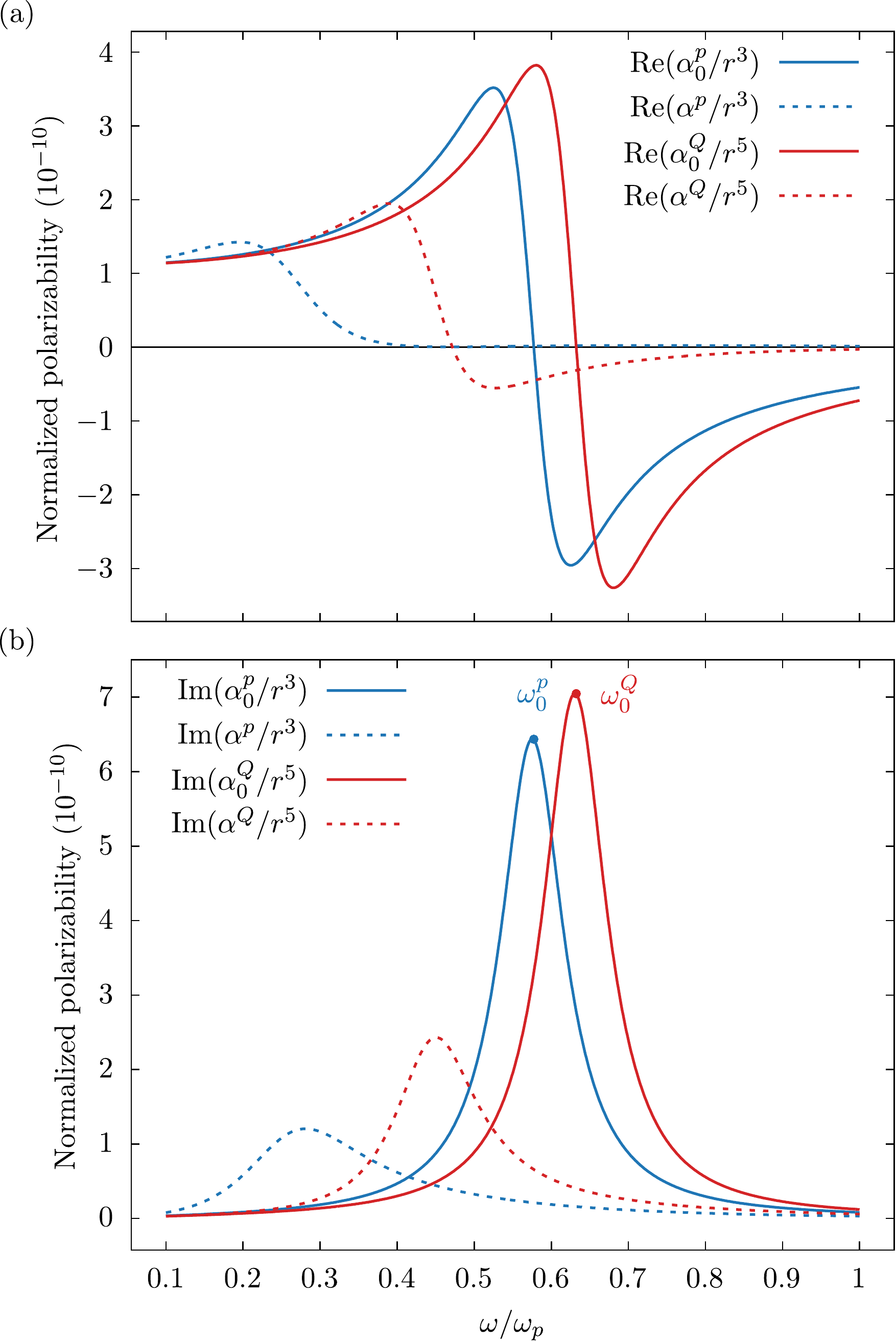}
    \caption{\label{fig2}Comparison of the electrostatic polarizabilities, $\alpha_0^p$ and $\alpha_0^Q$, with the polarizabilities from the Mie theory, $\alpha^p$ and $\alpha^Q$.
    (a) Real part of the normalized polarizability.
    (b) Imaginary part of the normalized polarizability.
    The results are calculated with $r=\SI{100}{\nano\metre}$ and $\gamma=0.1\omega_p$.
    }
\end{figure}
Strong dipole--quadrupole coupling can be realized in plasmonic meta-atoms such as H-like nanostructures\cite{goncalves2014}, T-shaped heterodimers\cite{guo2018}, and nanorod dimer\cite{pang2021}.
For simplicity, we consider homogeneous spherical nanoparticles with radius $r$.
We assume the permittivity of the nanoparticle is described by the Drude model
\begin{equation}
    \frac{\varepsilon(\omega)}{\varepsilon_0}=1-\frac{\omega_p^2}{\omega(\omega+i\gamma)},\label{eqiii1}
\end{equation}
where $\omega_p$ is the plasma frequency and $\gamma$ is the electron scattering rate.
The permeability of the nanoparticle is taken to be the same as the surrounding medium $\mu=\mu_0$.
The scattering of an electromagnetic plane wave by a homogeneous sphere can be obtained from the Mie theory.
The electric dipole polarizability
\begin{equation}
    \alpha^p=i4\pi\varepsilon_0\frac{3}{2k_0^3}a_1,\label{eqiii2}
\end{equation}
and the electric quadrupole polarizability\cite{babicheva2018,babicheva2019}
\begin{equation}
    \alpha^Q=i4\pi\varepsilon_0\frac{30}{k_0^5}a_2,\label{eqiii3}
\end{equation}
give the response of the nanoparticle to an electromagnetic field, where $a_n$ are the scattering coefficients given as
\begin{equation}
    a_n=\frac{m\psi_n(mx)\psi'_n(x)-\psi_n(x)\psi'_n(mx)}{m\psi_n(mx)\xi'_n(x)-\xi_n(x)\psi'_n(mx)},\label{eqiii4}
\end{equation}
in which $\psi_n$ and $\xi_n$ are the Riccati-Bessel functions, $x=k_0r$ is the size parameter, and $m=\sqrt{\varepsilon(\omega)/\varepsilon_0}$ is the relative refractive index.
We consider the power series expansion of the scattering coefficients to terms of order $x^6$\cite{bohren1998}
\begin{eqnarray}
    a_1&=&-\frac{i2x^3}{3}\frac{m^2-1}{m^2+2}-\frac{i2x^5}{5}\frac{(m^2-2)(m^2-1)}{(m^2+2)^2}\nonumber
    \\*
    &&+\frac{4x^6}{9}\left(\frac{m^2-1}{m^2+2}\right)^2+O(x^7),\label{eqiii5}
\end{eqnarray}
and
\begin{equation}
    a_2=-\frac{ix^5}{15}\frac{m^2-1}{2m^2+3}+O(x^7).\label{eqiii6}
\end{equation}
For sphere small compared with the wavelength ($x\ll1$, $\abs{m}x\ll1$), we get the approximate expressions by retaining the first term in each of the expansions.
Then we obtain the electrostatic dipole polarizability
\begin{equation}
    \alpha^p_0=4\pi\varepsilon_0r^3\frac{\varepsilon(\omega)-\varepsilon_0}{\varepsilon(\omega)+2\varepsilon_0},\label{eqiii7}
\end{equation}
and the electrostatic quadrupole polarizability
\begin{equation}
    \alpha^Q_0=4\pi\varepsilon_0r^5\frac{\varepsilon(\omega)-\varepsilon_0}{\varepsilon(\omega)+\frac{3}{2}\varepsilon_0}.\label{eqiii8}
\end{equation}

The dipole resonant frequency $\omega_0^p$ and the quadrupole resonant frequency $\omega_0^Q$ of the nanoparticle can be found by solving $\Re[\alpha^p(\omega_0^p)^{-1}]=0$ and $\Re[\alpha^Q(\omega_0^Q)^{-1}]=0$, respectively.
From the electrostatic polarizabilities, we find $\omega_0^p=\omega_p/\sqrt{3}$ and $\omega_0^Q=\omega_p\sqrt{2/5}$.
To compare the electrostatic polarizability with those from the Mie theory, the normalized polarizability $\alpha^p/r^3$ and $\alpha^Q/r^5$ are plotted in Fig.~\ref{fig2}.
The resonant frequencies can also be found from the peaks of $\Im(\alpha^p/r^3)$ and $\Im(\alpha^Q/r^5)$ in Fig.~\ref{fig2}(b).
We find that the electrostatic approximation introduces a blueshift to the resonant frequencies.

\section{\label{seciv}Analytical solutions of 1-D monopartite lattices}
\begin{figure}
    \includegraphics[width=8.6cm]{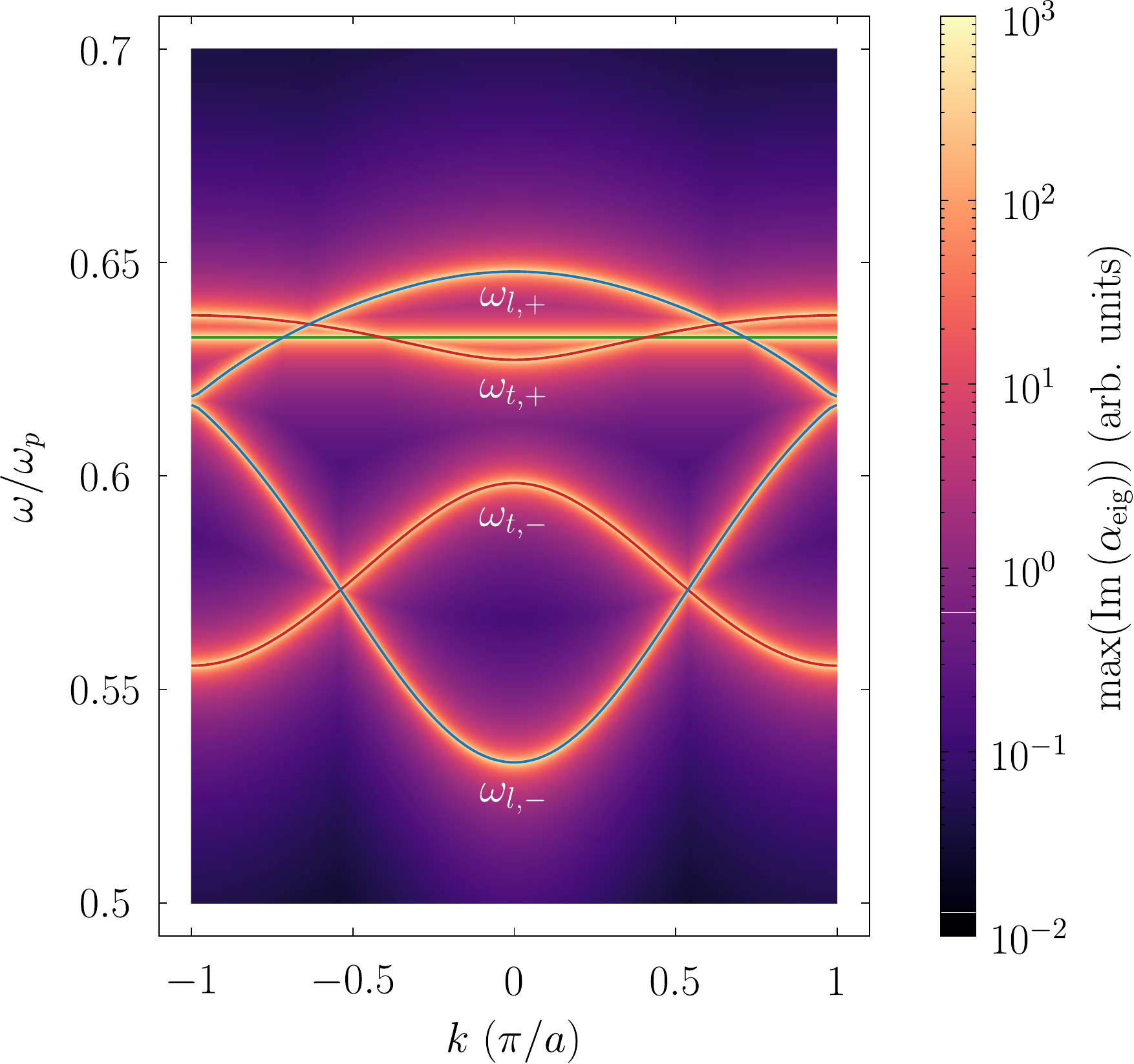}
    \caption{\label{fig3}Dispersion relations of 1-D infinite monopartite lattice of nanoparticles.
    The longitudinal modes $\omega_{l,\pm}(k)$ are plotted in blue and the transverse modes $\omega_{t,\pm}(k)$ are plotted in red.
    The localized quadrupole modes $\omega_{Q}=\omega_p\sqrt{2/5}$ are plotted in green.
    The analytical solutions are calculated with $\gamma=0$, while the numerical results with eigenresponse theory are calculated with $\gamma=0.001\omega_p$.
    The peaks of $\max(\Im{(\alpha_{\text{eig}})})$ represent resonances of the eigenmodes..
    }
\end{figure}
We consider 1-D infinite periodic monopartite lattice of nanoparticles.
The system is depicted in Fig.~\ref{fig1}.
The unit cell consist of one nanoparticle with radius $r$.
The position vector is given by $\bm{R}=na\hat{\bm{x}}$, where $a=3r$ is the lattice constant and $n$ is an integer.
The spectral properties of the systems are scale invariance which only depend on $r/a$ and we assume the nanoparticles have significant quadrupole response.
To obtain the dispersion relations, we consider there is no external excitation field such that $\bm{F}=0$.
Then the longitudinal modes are given by
\begin{widetext}
    \begin{equation}
        \left[
            \begin{pmatrix}
                (\alpha^p)^{-1}&0\\
                0&(\alpha^Q)^{-1}
            \end{pmatrix}
            -\frac{k_0^2}{\varepsilon_0}\sum_{\bm{R}\neq0}e^{i\bm{k}\cdot\bm{R}}
            \begin{pmatrix}
                G^p_{xx}&G^Q_{xx}n_x\\
                \frac{\partial G^p_{xx}}{\partial x}&\frac{\partial G^Q_{xx}n_x}{\partial x}
            \end{pmatrix}
            \right]
            \begin{pmatrix}
                p_x(\bm{r})\\
                Q_{xx}(\bm{r})
            \end{pmatrix}
            =\bm{0},\label{eqiv1}\\    
    \end{equation}
    and the transverse modes are given by
    \begin{equation}
        \left[
            \begin{pmatrix}
                (\alpha^p)^{-1}&0\\
                0&(\alpha^Q)^{-1}
            \end{pmatrix}
            -\frac{k_0^2}{\varepsilon_0}\sum_{\bm{R}\neq0}e^{i\bm{k}\cdot\bm{R}}
            \begin{pmatrix}
                G^p_{yy}&G^Q_{yy}n_x\\
                \frac{1}{2}\frac{\partial G^p_{yy}}{\partial x}&\frac{1}{2}\frac{\partial G^Q_{yy}n_x}{\partial x}
            \end{pmatrix}
            \right]
            \begin{pmatrix}
                p_y(\bm{r})\\
                Q_{xy}(\bm{r})
            \end{pmatrix}
            =\bm{0},\label{eqiv2}    
    \end{equation}
\end{widetext}
where the Bloch wave vector is given by $\bm{k}=k\hat{\bm{x}}$.
The transverse modes are degenerated with
\begin{equation}
    \begin{pmatrix}
        p_y(\bm{r})\\
        Q_{xy}(\bm{r})
    \end{pmatrix}
    =
    \begin{pmatrix}
        p_z(\bm{r})\\
        Q_{xz}(\bm{r})
    \end{pmatrix}.
\end{equation}
In addition, there are localized quadrupole modes
\begin{equation}
    \begin{pmatrix}
        (\alpha^Q)^{-1}&0\\
        0&(\alpha^Q)^{-1}
    \end{pmatrix}
    \begin{pmatrix}
        Q_{yy}(\bm{r})\\
        Q_{yz}(\bm{r})
    \end{pmatrix}
    =\bm{0},\label{eqiv3}
\end{equation}
which exist only in 1-D lattices.

In the quasi-electrostatic limit, we take the nearest neighbor approximation, Eq.~(\ref{eqiv1}) and Eq.~(\ref{eqiv2}) become
\begin{widetext}
    \begin{equation}
        \begin{pmatrix}
            \frac{1}{r^3}\left(1-3\frac{\omega^2}{\omega_p^2}\right)-\frac{4}{a^3}\cos(ka)&i\frac{3}{a^4}\sin(ka)\\
            -i\frac{12}{a^4}\sin(ka)&\frac{1}{r^5}\left(1-\frac{5}{2}\frac{\omega^2}{\omega_p^2}\right)+\frac{12}{a^5}\cos(ka)
        \end{pmatrix}
        \begin{pmatrix}
            p_x(\bm{r})\\
            Q_{xx}(\bm{r})    
        \end{pmatrix}
        =\bm{0},\label{eqiv4}
    \end{equation}
    and
    \begin{equation}
        \begin{pmatrix}
            \frac{1}{r^3}\left(1-3\frac{\omega^2}{\omega_p^2}\right)+\frac{2}{a^3}\cos(ka)&-i\frac{2}{a^4}\sin(ka)\\
            i\frac{3}{a^4}\sin(ka)&\frac{1}{r^5}\left(1-\frac{5}{2}\frac{\omega^2}{\omega_p^2}\right)-\frac{4}{a^5}\cos(ka)
        \end{pmatrix}
        \begin{pmatrix}
            p_y(\bm{r})\\
            Q_{xy}(\bm{r})    
        \end{pmatrix}
        =\bm{0}.\label{eqiv5}
    \end{equation}
\end{widetext}
After solving, the dispersion relations for the longitudinal modes are
\begin{equation}
    \omega_{l,\pm}(k)=\omega_p\sqrt{f_{l,b}(k)\pm\sqrt{f_{l,b}^2(k)-f_{l,c}(k)}},\label{eqiv6}\\
\end{equation}
and the dispersion relations for the transverse modes are
\begin{equation}
    \omega_{t,\pm}(k)=\omega_p\sqrt{f_{t,b}(k)\pm\sqrt{f_{t,b}^2(k)-f_{t,c}(k)}},\label{eqiv7}\\
\end{equation}
where
\begin{widetext}
    \begin{subequations}
        \label{eqiv8}
        \begin{eqnarray}
            f_{l,b}(k)&:=&-\frac{1}{2}\left[-\frac{11}{15}+\frac{4}{3}\left(\frac{r}{a}\right)^3\cos(ka)-\frac{24}{5}\left(\frac{r}{a}\right)^5\cos(ka)\right],\label{eqiv8a}
            \\
            f_{l,c}(k)&:=&\frac{2}{15}-\frac{8}{15}\left(\frac{r}{a}\right)^3\cos(ka)+\frac{8}{5}\left(\frac{r}{a}\right)^5\cos(ka)-\frac{32}{5}\left(\frac{r}{a}\right)^8\cos^2(ka)-\frac{24}{5}\left(\frac{r}{a}\right)^8\sin^2(ka),\label{eqiv8b}
            \\
            f_{t,b}(k)&:=&-\frac{1}{2}\left[-\frac{11}{15}-\frac{2}{3}\left(\frac{r}{a}\right)^3\cos(ka)+\frac{8}{5}\left(\frac{r}{a}\right)^5\cos(ka)\right],\label{eqiv8c}
            \\
            f_{t,c}(k)&:=&\frac{2}{15}+\frac{4}{15}\left(\frac{r}{a}\right)^3\cos(ka)-\frac{8}{15}\left(\frac{r}{a}\right)^5\cos(ka)-\frac{16}{15}\left(\frac{r}{a}\right)^8\cos^2(ka)-\frac{4}{5}\left(\frac{r}{a}\right)^8\sin^2(ka).\label{eqiv8d}
        \end{eqnarray}
    \end{subequations}
\end{widetext}
The eigenmodes of the longitudinal modes read
\begin{equation}
    \begin{pmatrix}
        p_x(\bm{r})\\
        Q_{xx}(\bm{r})    
    \end{pmatrix}
    =\frac{1}{\sqrt{1+A_{l,\pm}^2(k)}}
    \begin{pmatrix}
        1\\
        A_{l,\pm}(k)e^{i\frac{\pi}{2}}
    \end{pmatrix},\label{eqiv9}\\
\end{equation}
and the eigenmodes of the transverse modes read
\begin{equation}
    \begin{pmatrix}
        p_y(\bm{r})\\
        Q_{xy}(\bm{r})    
    \end{pmatrix}
    =\frac{1}{\sqrt{1+A_{t,\pm}^2(k)}}
    \begin{pmatrix}
        1\\
        A_{t,\pm}(k)e^{-i\frac{\pi}{2}}
    \end{pmatrix},\label{eqiv10}
\end{equation}
where
\begin{widetext}
    \begin{subequations}
        \label{eqiv11}
        \begin{eqnarray}
            A_{l,\pm}&:=&\frac{a}{3\sin(ka)}\left[\left(\frac{a}{r}\right)^3\left[1-3\left(f_{l,b}\pm\sqrt{f_{l,b}^2-f_{l,c}}\right))\right]-4\cos(ka)\right],\label{eqiv11a}
            \\
            A_{t,\pm}&:=&\frac{a}{2\sin(ka)}\left[\left(\frac{a}{r}\right)^3\left[1-3\left(f_{t,b}\pm\sqrt{f_{t,b}^2-f_{t,c}}\right))\right]+2\cos(ka)\right].\label{eqiv11b}
        \end{eqnarray}
    \end{subequations}
\end{widetext}
We see that the dipole moments and the quadrupole moments have $\pi/2$ phase difference in both longitudinal modes and transverse modes.
In longitudinal modes, the dipole moment $p_x$ leads the quadrupole moment $Q_{xx}$, while in transverse mode, the dipole moment $p_y$ lags behind the quadrupole moment $Q_{xy}$.
The localized quadrupole modes of Eq.~(\ref{eqiv3}) give a flat band at the quadrupole resonant frequency of the nanoparticle
\begin{equation}
    \omega_Q=\omega_p\sqrt{\frac{2}{5}},
\end{equation}
which is independent of $k$ with quadrupole moments
\begin{subequations}
    \label{eqiv12}
    \begin{eqnarray}
        Q_{yy}(\bm{r})&=&1,\label{eqiv12a}
        \\
        Q_{yz}(\bm{r})&=&1.\label{eqiv12b}
    \end{eqnarray}
\end{subequations}
All the dispersion relations are plotted in Fig.~\ref{fig3}.

The two longitudinal bands with solutions $\bm{X}_{l,\pm}(k)=(p_{x,\pm}(k),0,0,Q_{xx,\pm}(k),0,0,0,0)^T$ give $\bm{p}_{l,\pm}(k)=(p_{x,\pm}(k),0,0)^T$ and
\begin{equation}
    \bm{Q}_{l,\pm}(k)=
    \begin{pmatrix}
        Q_{xx,\pm}(k)&0&0\\
        0&0&0\\
        0&0&-Q_{xx,\pm}(k)
    \end{pmatrix},\label{eqiv13}
\end{equation}
whereas the two transverse bands with solutions $\bm{X}_{t,\pm}(k)=(0,p_{y,\pm}(k),0,0,Q_{xy,\pm}(k),0,0,0)^T$ give $\bm{p}_{t,\pm}(k)=(0,p_{y,\pm}(k),0)^T$ and
\begin{equation}
    \bm{Q}_{t,\pm}(k)=
    \begin{pmatrix}
        0&Q_{xy,\pm}(k)&0\\
        Q_{xy,\pm}(k)&0&0\\
        0&0&0
    \end{pmatrix}.\label{eqiv14}
\end{equation}
At the zone center $k=0$, the longitudinal eigenmodes read
\begin{subequations}
    \label{eqiv15}
    \begin{eqnarray}
        \begin{pmatrix}
            p_{x,-}\\
            Q_{xx,-}
        \end{pmatrix}
        &=&
        \begin{pmatrix}
            1\\
            0
        \end{pmatrix},\label{eqiv15a}
        \\
        \begin{pmatrix}
            p_{x,+}\\
            Q_{xx,+}
        \end{pmatrix}
        &=&
        \begin{pmatrix}
            0\\
            1
        \end{pmatrix},\label{eqiv15b}
    \end{eqnarray}
\end{subequations}
such that the lower longitudinal band is dipole dominated and the upper longitudinal band is quadrupole dominated.
Similarly, the transverse eigenmodes read
\begin{subequations}
    \label{eqiv16}
    \begin{eqnarray}
        \begin{pmatrix}
            p_{y,-}\\
            Q_{xy,-}
        \end{pmatrix}
        &=&
        \begin{pmatrix}
            1\\
            0
        \end{pmatrix},\label{eqiv16a}
        \\
        \begin{pmatrix}
            p_{y,+}\\
            Q_{xy,+}
        \end{pmatrix}
        &=&
        \begin{pmatrix}
            0\\
            1
        \end{pmatrix},\label{eqiv16b}
    \end{eqnarray}
\end{subequations}
such that the lower transverse band is dipole dominated and the upper transverse band is quadrupole dominated.
On the other hand, at the zone boundary $k=\pi/a$, the longitudinal eigenmodes read
\begin{subequations}
    \label{eqiv17}
    \begin{eqnarray}
        \begin{pmatrix}
            p_{x,-}\\
            Q_{xx,-}
        \end{pmatrix}
        &=&
        \begin{pmatrix}
            0\\
            1
        \end{pmatrix},\label{eqiv17a}
        \\
        \begin{pmatrix}
            p_{x,+}\\
            Q_{xx,+}
        \end{pmatrix}
        &=&
        \begin{pmatrix}
            1\\
            0
        \end{pmatrix},\label{eqiv17b}
    \end{eqnarray}
\end{subequations}
such that the lower longitudinal band is quadrupole dominated and the upper longitudinal band is dipole dominated.
In contrast, the transverse eigenmodes remain unchanged with
\begin{subequations}
    \label{eqiv18}
    \begin{eqnarray}
        \begin{pmatrix}
            p_{y,-}\\
            Q_{xy,-}
        \end{pmatrix}
        &=&
        \begin{pmatrix}
            1\\
            0
        \end{pmatrix},\label{eqiv18a}
        \\
        \begin{pmatrix}
            p_{y,+}\\
            Q_{xy,+}
        \end{pmatrix}
        &=&
        \begin{pmatrix}
            0\\
            1
        \end{pmatrix}.\label{eqiv18b}
    \end{eqnarray}
\end{subequations}
The quadrupole bands with solutions $\bm{X}_{Q}=(0,0,0,0,0,0,Q_{yy},Q_{yz})^T$ yield $\bm{p}_Q=(0,0,0)^T$ and
\begin{equation}
    \bm{Q}_{Q}=
    \begin{pmatrix}
        0&0&0\\
        0&Q_{yy}&Q_{yz}\\
        0&Q_{yz}&-Q_{yy}
    \end{pmatrix}.\label{eqiv19}
\end{equation}
We see that $\bm{X}_{l,\pm}(k)$, $\bm{X}_{t,\pm}(k)$, and $\bm{X}_{Q}$ are orthogonal to each other.

Previous works on plasmonic nanoparticles in 1-D lattices are limited to either dipole--dipole interactions\cite{weber2004} or quadrupole--quadrupole interactions\cite{alu2009} such that band structures with only pure dipolar modes or pure quadrupolar modes are studied.
Our results extend those works by including all dipole--dipole, quadrupole--quadrupole, and dipole--quadrupole interactions, which cover all the bands presented in previous works and in addition with an extra quadrupolar flat band.
Besides dipole moments, this also reveal the contribution of quadrupole moments to the near-field interactions.

\section{\label{secv}Infinite bipartite lattice}
\begin{figure}
    \includegraphics[width=8.6cm]{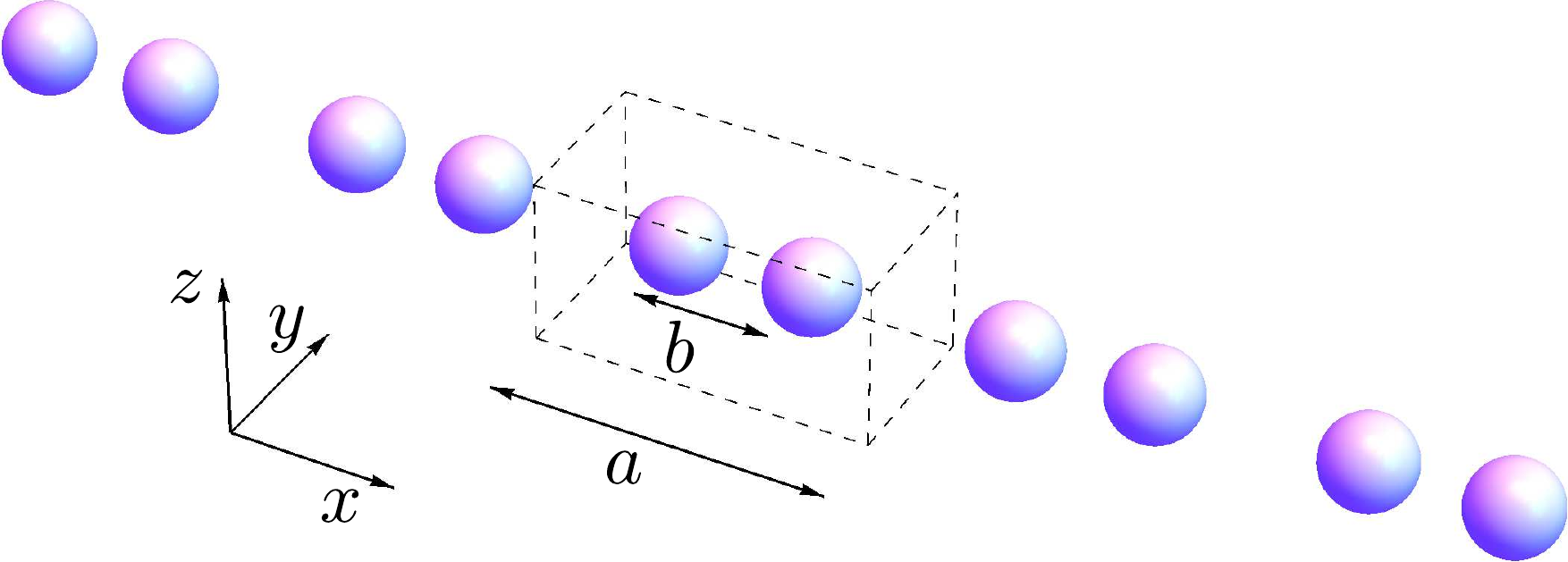}
    \caption{\label{fig4}Illustration of a 1-D bipartite lattice.
    The unit cell is indicated by the dashed box.
    The lattice constant is given by $a$ and the distance between nanoparticle $A$ and nanoparticle $B$ is given by $b=(1-\delta)a/2$.
    }
\end{figure}
\begin{figure*}
    \includegraphics[width=17.8cm]{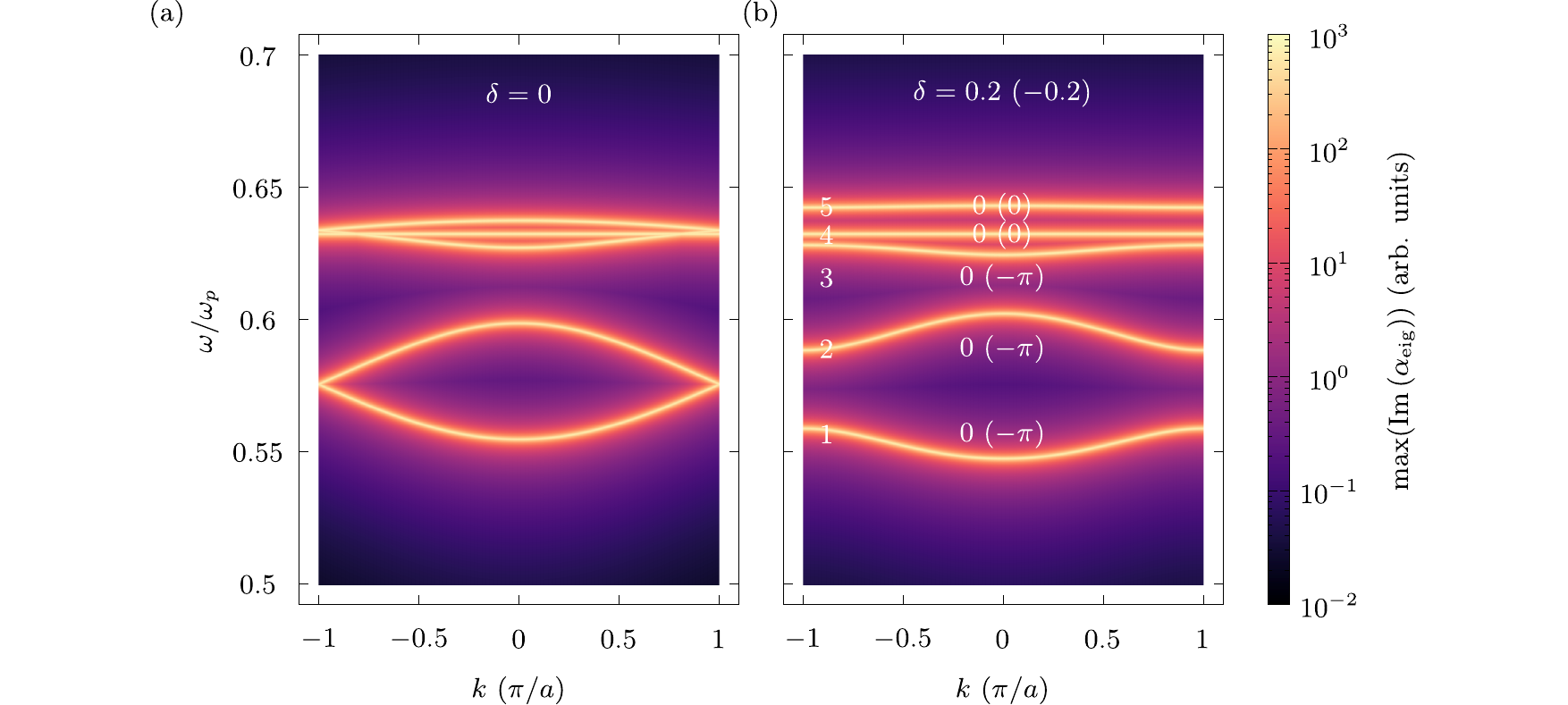}
    \caption{\label{fig5}Band structure of 1-D infinite bipartite lattice of nanoparticles with (a) $\delta=0$ and (b) $\delta=0.2$.
    The peaks of $\max(\Im{(\alpha_{\text{eig}})})$ represent resonances of the eigenmodes.
    Only longitudinal modes are shown.
    The Zak phase $\theta$ for the bands with $\delta<0$ ($\delta>0$) are labelled with corresponding band index $n$.
    The results are calculated with $\gamma=0.001\omega_p$.
    }
\end{figure*}
\begin{figure*}
    \includegraphics[width=17.8cm]{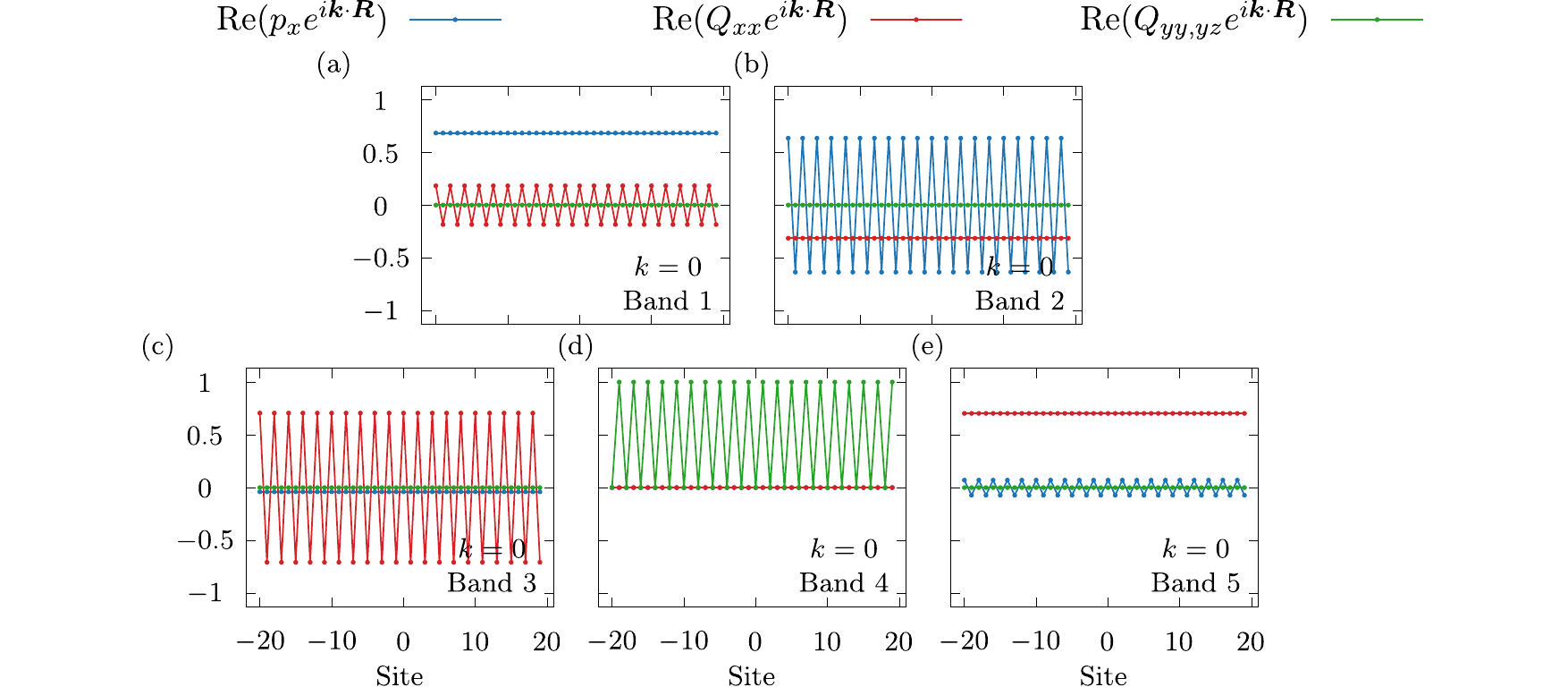}
    \caption{\label{fig6}The longitudinal eigenmodes of the 1-D infinite periodic bipartite lattice of nanoparticles with $\delta=0.2$ at the Brillouin zone center $k=0$ corresponding to the band structures in Fig.~\ref{fig5}(b).
    $20$ unit cells are shown.
    }
\end{figure*}
\begin{figure*}
    \includegraphics[width=17.8cm]{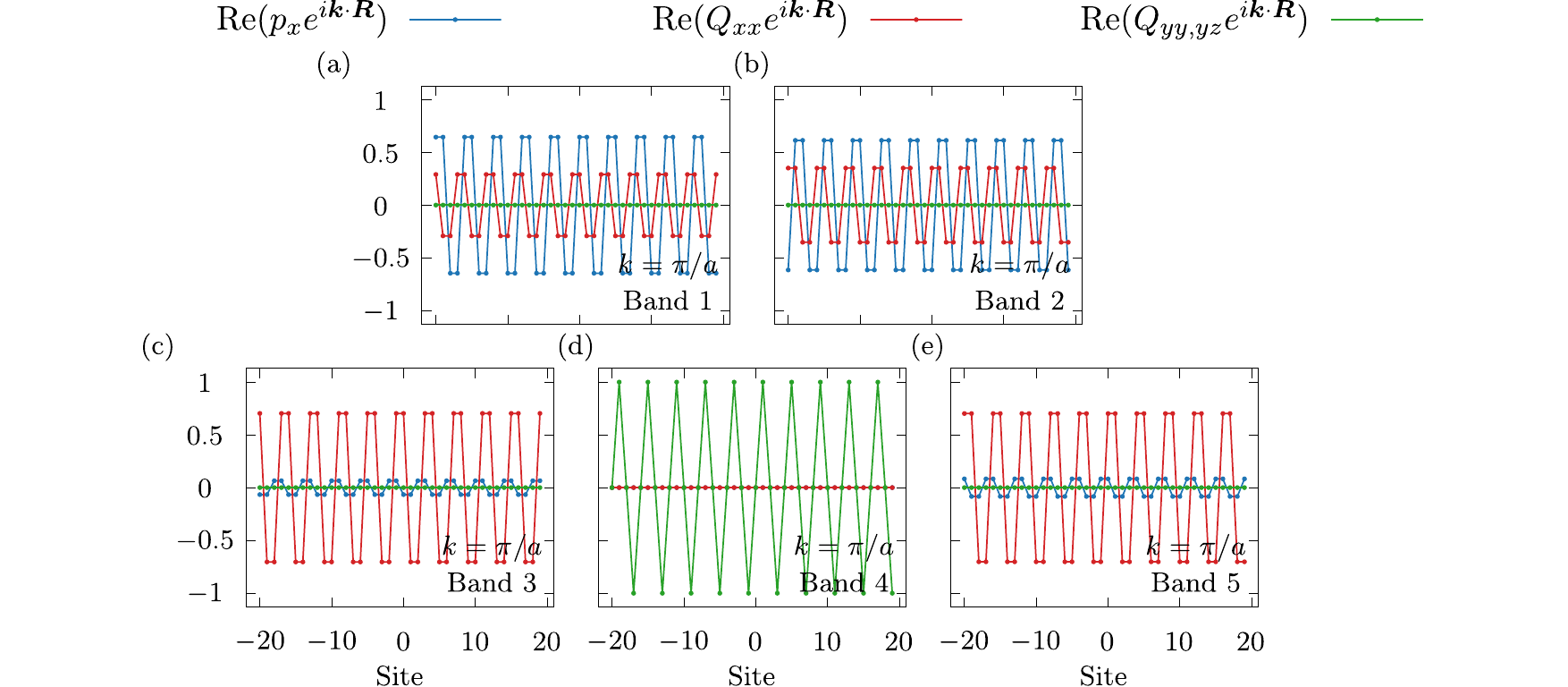}
    \caption{\label{fig7}The longitudinal eigenmodes of the 1-D infinite periodic bipartite lattice of nanoparticles with $\delta=0.2$ at the Brillouin zone boundary $k=\pi/a$ corresponding to the band structures in Fig.~\ref{fig5}(b).
    $20$ unit cells are shown.
    }
\end{figure*}
\begin{figure*}
    \includegraphics[width=17.8cm]{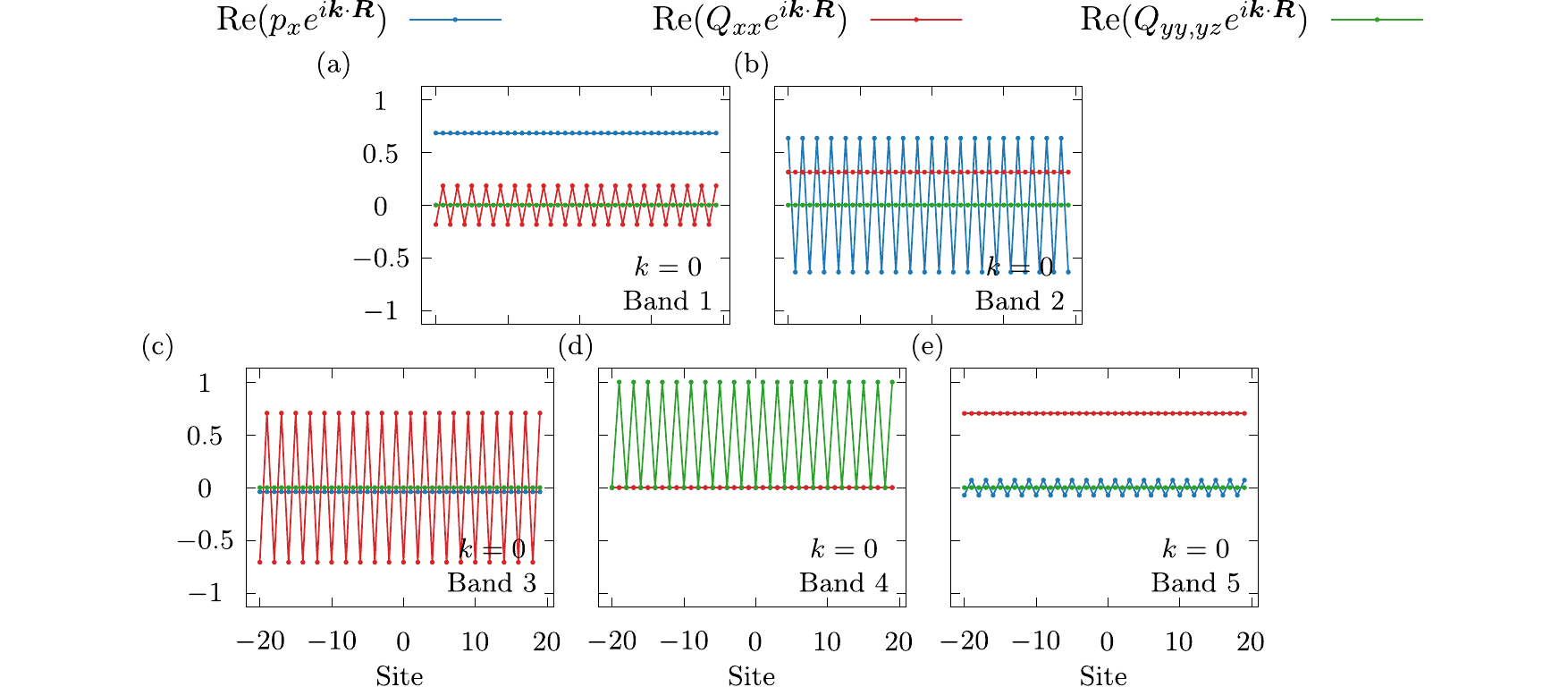}
    \caption{\label{fig8}The longitudinal eigenmodes of the 1-D infinite periodic bipartite lattice of nanoparticles with $\delta=-0.2$ at the Brillouin zone center $k=0$ corresponding to the band structures in Fig.~\ref{fig5}(b).
    $20$ unit cells are shown.
    }
\end{figure*}
\begin{figure*}
    \includegraphics[width=17.8cm]{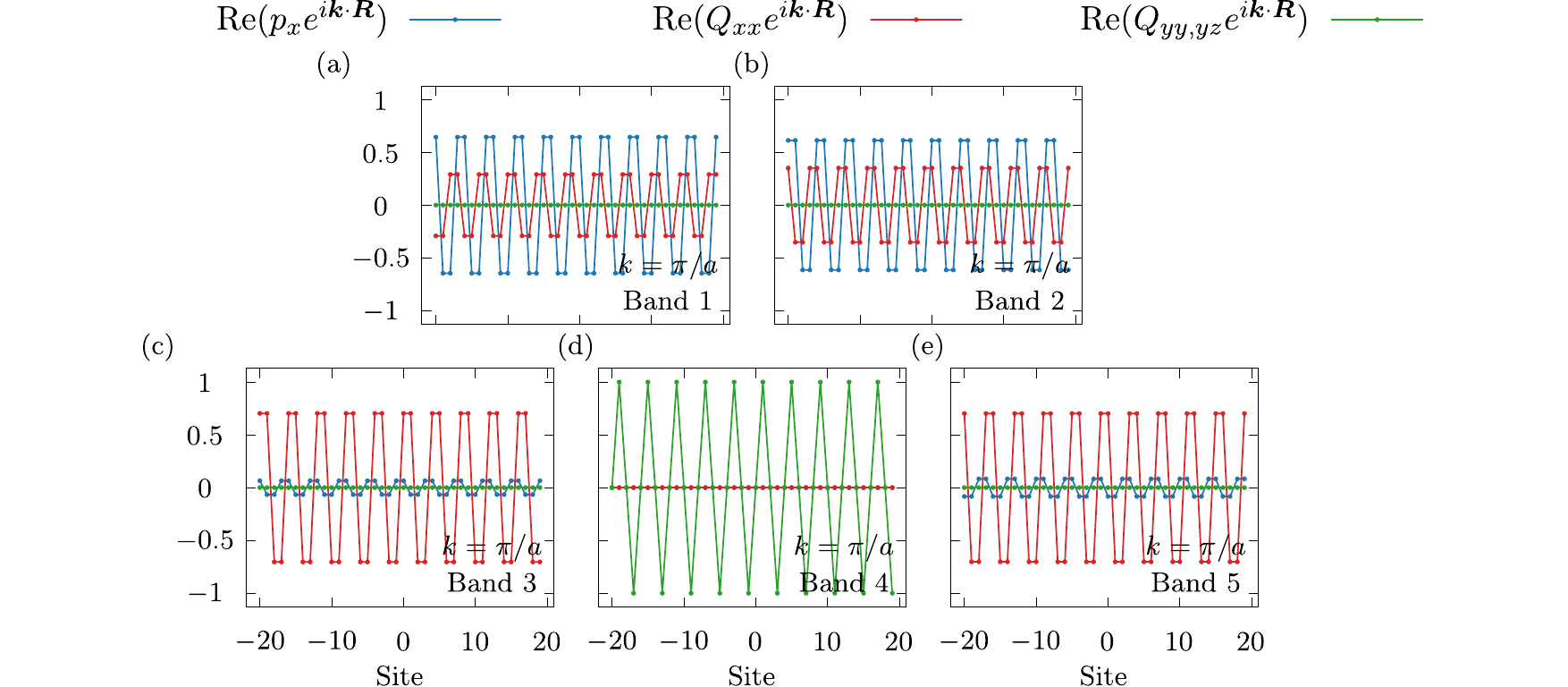}
    \caption{\label{fig9}The longitudinal eigenmodes of the 1-D infinite periodic bipartite lattice of nanoparticles with $\delta=-0.2$ at the Brillouin zone boundary $k=\pi/a$ corresponding to the band structures in Fig.~\ref{fig5}(b).
    $20$ unit cells are shown.
    }
\end{figure*}
We now limit our scope in the longitudinal modes and the localized quadrupole modes of a bipartite model as depicted in Fig.~\ref{fig4}.
The unit cell consists of two nanoparticles, labeled as $A$ and $B$.
The displacement from nanoparticle $A$ to nanoparticle $B$ is given by $\bm{b}=b\hat{\bm{x}}$, with
\begin{equation}
    b=\frac{a}{2}(1-\delta),\label{eqv1}
\end{equation}
where $\delta$ is a dimensionless parameter.
For $\delta=0$, the nanoparticles are in equidistance as depicted in Fig.~\ref{fig1}, and its band structure given in Fig.~\ref{fig3}.
For any $\delta\neq0$, the lattices are dimerized.
For the longitudinal modes, the state vectors of nanoparticles $A$ and nanoparticles $B$ are given by $\bm{X}_A=(p_{A,x},Q_{A,xx})^T$ and $\bm{X}_B=(p_{B,x},Q_{B,xx})^T$, respectively.
The polarizabilities of nanoparticles $A$ and nanoparticles $B$ are given by $\bm{\alpha}_A=\diag(\alpha_A^p,\alpha_A^Q)$ and $\bm{\alpha}_B=\diag(\alpha_B^p,\alpha_B^Q)$, respectively.
The coupled dipole--quadrupole equations for the bipartite model are then formulated as
\begin{widetext}
    \begin{equation}
        \begin{pmatrix}
            \bm{\alpha}_A^{-1}&\bm{0}\\
            \bm{0}&\bm{\alpha}_B^{-1}
        \end{pmatrix}
        \begin{pmatrix}
            \bm{X}_A\\
            \bm{X}_B
        \end{pmatrix}
        =\sum_{\bm{R}\neq0}e^{i\bm{k}\cdot\bm{R}}
        \begin{pmatrix}
            \bm{\Gamma}^{AA}(0,\bm{R})&\bm{\Gamma}^{AB}(0,\bm{R}+\bm{b})\\
            \bm{\Gamma}^{BA}(0,\bm{R}-\bm{b})&\bm{\Gamma}^{BB}(0,\bm{R})
        \end{pmatrix}
        \begin{pmatrix}
            \bm{X}_A\\
            \bm{X}_B
        \end{pmatrix}.\label{eqv2}
    \end{equation}
\end{widetext}
In the quasi-electrostatic limit with nearest neighbor approximation, we have explicitly,
\begin{widetext}
    \begin{equation}
        \frac{1}{4\pi\varepsilon_0}
        \begin{pmatrix}
            \alpha_A^p&0&-\frac{2}{b^3}-\frac{2}{(a-b)^3}e^{-ika}&\frac{3}{2b^4}-\frac{3}{2(a-b)^4}e^{-ika}\\
            0&\alpha_A^Q&-\frac{6}{b^4}+\frac{6}{(a-b)^4}e^{-ika}&\frac{6}{b^5}+\frac{6}{(a-b)^5}e^{-ika}\\
            -\frac{2}{b^3}-\frac{2}{(a-b)^3}e^{ika}&-\frac{3}{2b^4}+\frac{3}{2(a-b)^4}e^{ika}&\alpha_B^p&0\\
            \frac{6}{b^4}-\frac{6}{(a-b)^4}e^{ika}&\frac{6}{b^5}+\frac{6}{(a-b)^5}e^{ika}&0&\alpha_B^Q
        \end{pmatrix}
        \begin{pmatrix}
            p_{A,x}\\
            Q_{A,xx}\\
            p_{B,x}\\
            Q_{B,xx}
        \end{pmatrix}
        =0.\label{eqv3}
    \end{equation}
\end{widetext}
In addition, the localized quadrupole modes are again given by Eq.~(\ref{eqiv3}).

Instead of solving Eq.~(\ref{eqv3}) directly, we use an eigenresponse theory\cite{bergman1980,markel1995,fung2007,fung2008} to study the spectral response of the system, which is based on spectral decomposition and has been extensively used for studying plasmonic\cite{zhang2018,zhang2020} and gyromagnetic lattices\cite{rphwu2019}.
In the eigenresponse theory, we consider the eigenvalue problem
\begin{equation}
    \bm{M}(k,\omega)\bm{X}_i=\lambda_i(k,\omega)\bm{X}_i,\label{eqv4}
\end{equation}
where we define
\begin{equation}
    \bm{M}(k,\omega):=\bm{A}(\omega)-\bm{\Gamma}(k),\label{eqv5}
\end{equation}
and $\lambda_i(k,\omega)$ is the eigenvalue corresponding to the eigenmode $\bm{X}_i$.
The eigenpolarizability
\begin{equation}
    \alpha_{\text{eig}}(k,\omega):=\frac{1}{\lambda_i(k,\omega)},\label{eqv6}
\end{equation}
can be interpreted as the response function of the corresponding eigenmode for an external excitation field and the peaks of $\Im(\alpha_{\text{eig}})$ represent resonances.
We solve Eq.~(\ref{eqiv4}) and Eq.~(\ref{eqiv5}) again with eigenresponse theory numerically to show the validity.
The results are shown in the colormap of Fig.~\ref{fig3}, in which the peaks of $\max(\Im(\alpha_{\text{eig}}))$ define the resonances of the eigenmodes.
We see that the numerical results agree with the analytical solutions.

For the bipartite model described by Eq.~(\ref{eqv3}), we consider three cases with different dimerization parameter, $\delta=0$ and $\delta=\pm0.2$.
For $\delta=0$, the system is the same as the one discussed in Sec.~\ref{seciv} and the corresponding band structure is shown in Fig.~\ref{fig5}(a).
Apart from the quadrupolar flat band, four bands are obtained for the longitudinal modes due to the band folding, and they are physically the same as those in Fig.~\ref{fig3}.
There is a band gap between two sets of bands, but we will soon see that it is topologically trivial.
Apart from that, the system is gapless as there are degeneracies at the zone boundary $k=\pm\pi/a$.
For $\delta=\pm0.2$, this corresponds to a different choice for the unit cell of the system.
Both band structures are the same as shown in Fig.~\ref{fig5}(b).
In fact, for any $\delta\neq0$, as the inversion symmetry of the system is now reduced, the degeneracies in Fig.~\ref{fig5}(a) at the zone boundary are removed resulting in a gap.
We now have five bands that are fully gapped.

The eigenmodes of the infinite bipartite lattice for $\delta=0.2$ at the zone center $k=0$ and the zone boundary $k=\pi/a$ are shown in Fig.~\ref{fig6} and Fig.~\ref{fig7}, respectively.
Also, the eigenmodes of the infinite bipartite lattice for $\delta=-0.2$ at the zone center $k=0$ and the zone boundary $k=\pi/a$ are shown in Fig.~\ref{fig8} and Fig.~\ref{fig9}, respectively.
We observe that, apart from Band 4, the dipole moments and the quadrupole moments within a unit cell always have different symmetries.
This can be explained by the analytical solution of monopartite lattice in Eq.~(\ref{eqiv9}) and (\ref{eqiv10}), where there is a $\pi/2$ phase difference in the dipole and quadrupole moments.
At the zone center $k=0$, for both $\delta=\pm0.2$, the dipole moments are in-phase in Band 1 and Band 3, and are anti-phase in Band 2 and Band 5, while the quadrupole moments are in-phase in Band 2 and Band 5, and are anti-phase in Band 1 and Band 3.
At the zone boundary $k=\pi/a$, the eigenmodes behave in the same way for $\delta=0.2$.
In contrast, for $\delta=-0.2$, the dipole moments are in-phase in Band 2 and Band 5, and are anti-phase in Band 1 and Band 3, while the quadrupole moments are in-phase in Band 1 and Band 3, and are anti-phase in Band 2 and Band 5.
In both cases, the quadrupole components $Q_{yy,yz}$ are non-zero only in Band 4, and they are in-phase at $k=0$ and are anti-phase at $k=\pi/a$.
Furthermore, we observe that Band 1 and Band 2 are dipole dominated such that the dipole moments have higher energy, while Band 3 and Band 5 are quadrupole dominated such that the quadrupole moments have higher energy.
It is because at high frequency, the oscillation of higher-order multipoles is favored.

\subsection{\label{secva}Topological phase transitions}
\begin{figure*}
    \includegraphics[width=17.8cm]{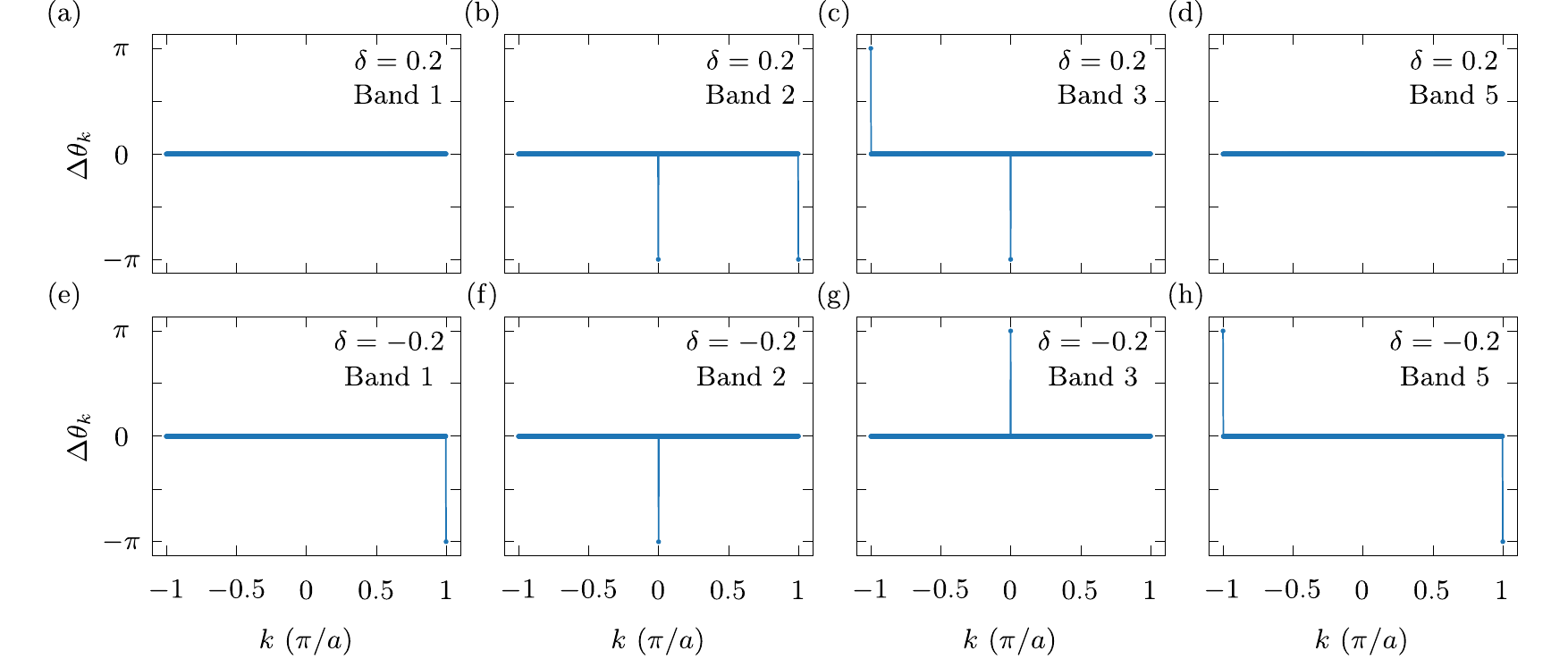}
    \caption{\label{fig10}Phase difference $\Delta\theta_k$ between the states $\bm{X}_{n,k}$ and $\bm{X}_{n,k+\Delta k}$ for the bands correspond to the band structure in Fig.~\ref{fig5}(b).
    The results are calculated from $k=-\pi/a$ to $k=\pi/a-\Delta k$ with $N=1000$.
    }
\end{figure*}
\begin{figure}
    \includegraphics[width=8.6cm]{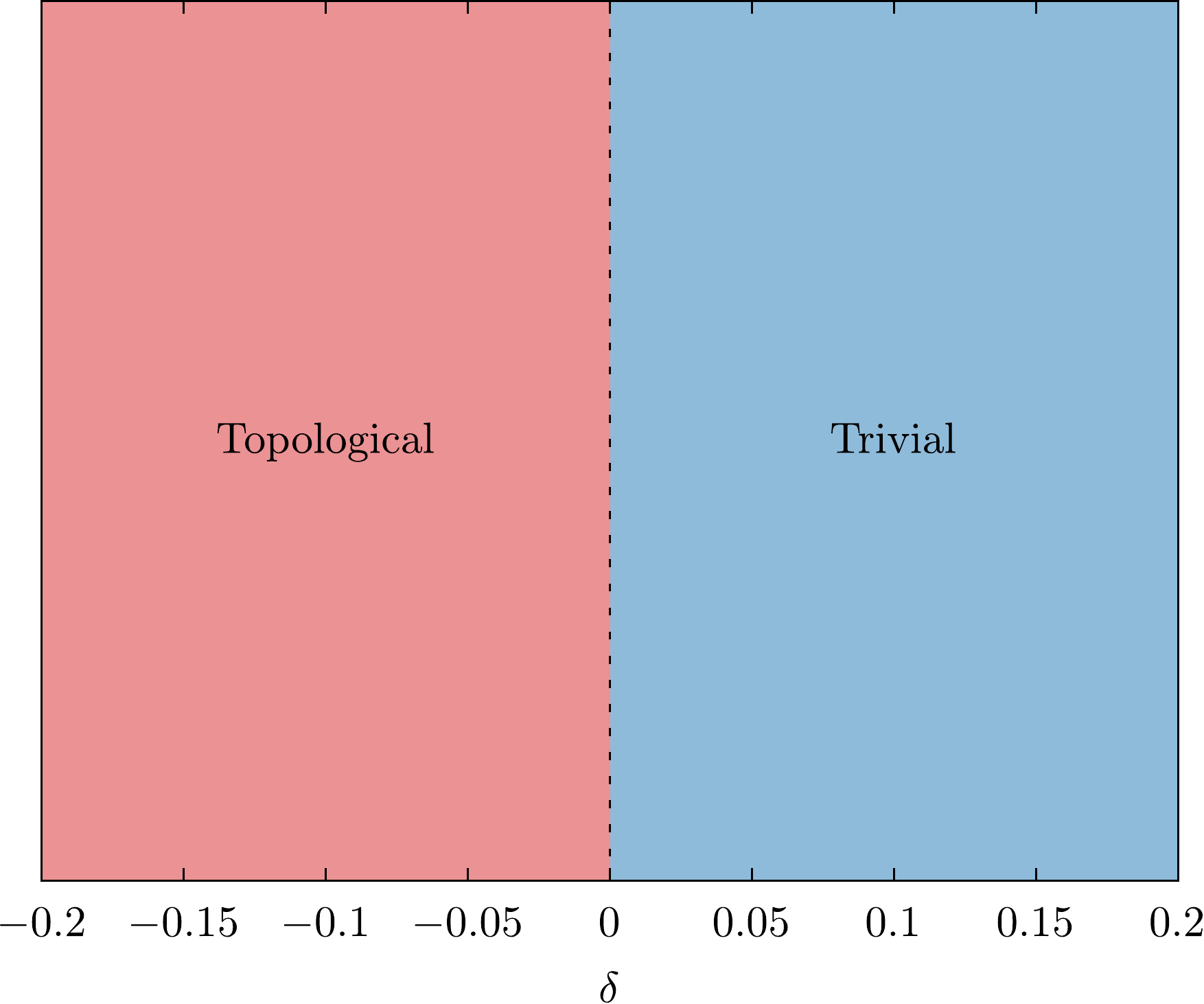}
    \caption{\label{fig11}Topological phase diagram of a 1-D bipartite lattice.
    The system is topologically trivial for $\delta>0$ and is topologically nontrivial for $\delta<0$.
    At $\delta=0$, the system undergoes topological phase transition.
    }
\end{figure}
We now classify the topology of the bands in Fig.~\ref{fig5}(b).
The topological invariant for 1-D system is given by the Zak phase\cite{zak1989}, it is defined as
\begin{equation}
    \theta_n=\oint_{\text{BZ}}\braket{\bm{X}_n|i\partial_k\bm{X}_n},\label{eqva1}
\end{equation}
where $n$ is the band index and $\braket{\bm{X}_n|i\partial_k\bm{X}_n}$ is also known as the Berry connection.
The Zak phase for system with inversion symmetry is a $Z_2$ invariant and quantized as $\theta=q\pi\mod(2\pi)$ with integer $q$.
We calculate the Zak phase for each bands numerically by discretizing the first Brillouin zone with $\Delta k=2\pi/Na$, where $N$ is the number of unit cell.
Then the Zak phase can be calculated using the Wilson loop approach\cite{asboth2016,wang2019}
\begin{equation}
    \theta_n=-\arg\left(\prod_k^{k-\Delta k}\frac{\braket{\bm{X}_{n,k}|\bm{X}_{n,k+\Delta k}}}{\norm{\braket{\bm{X}_{n,k}|\bm{X}_{n,k+\Delta k}}}}\right).\label{eqva2}
\end{equation}
It can also be expressed as
\begin{equation}
    \theta_n=\sum_{k}^{k-\Delta k}\Delta\theta_{n,k},\label{eqva3}
\end{equation}
where $\Delta\theta_{n,k}$ is the phase difference between the states $\bm{X}_{n,k}$ and $\bm{X}_{n,k+\Delta k}$.
In the continuum limit, $N\to\infty$ and $\Delta k\to0$, then Eq.~(\ref{eqva3}) recovers Eq.~(\ref{eqva1}).

Since the solution of the quadrupolar flat band is independent of $k$, Band 4 remains topological trivial for all $\delta$ with $\theta=0$.
We show the phase difference $\Delta\theta_{n,k}$ for each of the other bands in Fig.~\ref{fig10}.
The resulting Zak phases $\theta$ for all the bands are labeled in Fig.~\ref{fig5}(b).
We see that, for $\delta=0.2$, all bands have $\theta=0$, implying the system is topologically trivial.
On the other hand, for $\delta=-0.2$, we see while Band 1 to 3 have $\theta=-\pi$, are topologically non-trivial, Band 5 remains trivial with $\theta=0$.
We observe that all non-zero phase differences happens at either the zone center $k=0$ or the zone boundary $k=\pm\pi/a$.
In addition, the Zak phases are consistent with the field symmetry consideration\cite{xiao2014}.
It is known that the field symmetries at the Brillouin zone center and boundary are the same when $\theta=0$ but reversed when $\theta=\pi$.
Following the discussions in Sec.~\ref{secv}, for the bipartite lattice with $\delta=0.2$, we find from Fig.~\ref{fig6} and Fig.~\ref{fig7} that the dipolar and quadrupolar eigenmodes for all bands have the same symmetries at the zone center $k=0$ and the zone boundary $k=\pm\pi/a$, verifying $\theta=0$.
On the other hand, for the bipartite lattice with $\delta=-0.2$, we find from Fig.~\ref{fig8} and Fig.~\ref{fig9} that their eigenmode symmetries are different, giving $\theta=\pi$.

From the bulk-boundary correspondence, the existence of topological edge states depends on the summations of Zak phases below the gap\cite{xiao2014}.
If two systems with different summation of Zak phases below the $n$th gap are connected, it is expected that there is an edge state localized at the interface in the $n$th gap.
Then the band gap between Band 1 and Band 2 is topological.
Although for $\delta=-0.2$, Band 5 have trivial Zak phase $\theta=0$, the band gap between Band 3 and Band 5 is also topological, while the gap between Band 2 and Band 3 is trivial.
The topological phase diagram of a 1-D bipartite lattice is shown in Fig.~\ref{fig11}.
The system is topologically trivial for $\delta>0$ and is topologically nontrivial for $\delta<0$.
At $\delta=0$, the system undergoes topological phase transition.

\section{\label{secvi}Topological edge states}
\begin{figure}
    \includegraphics[width=8.6cm]{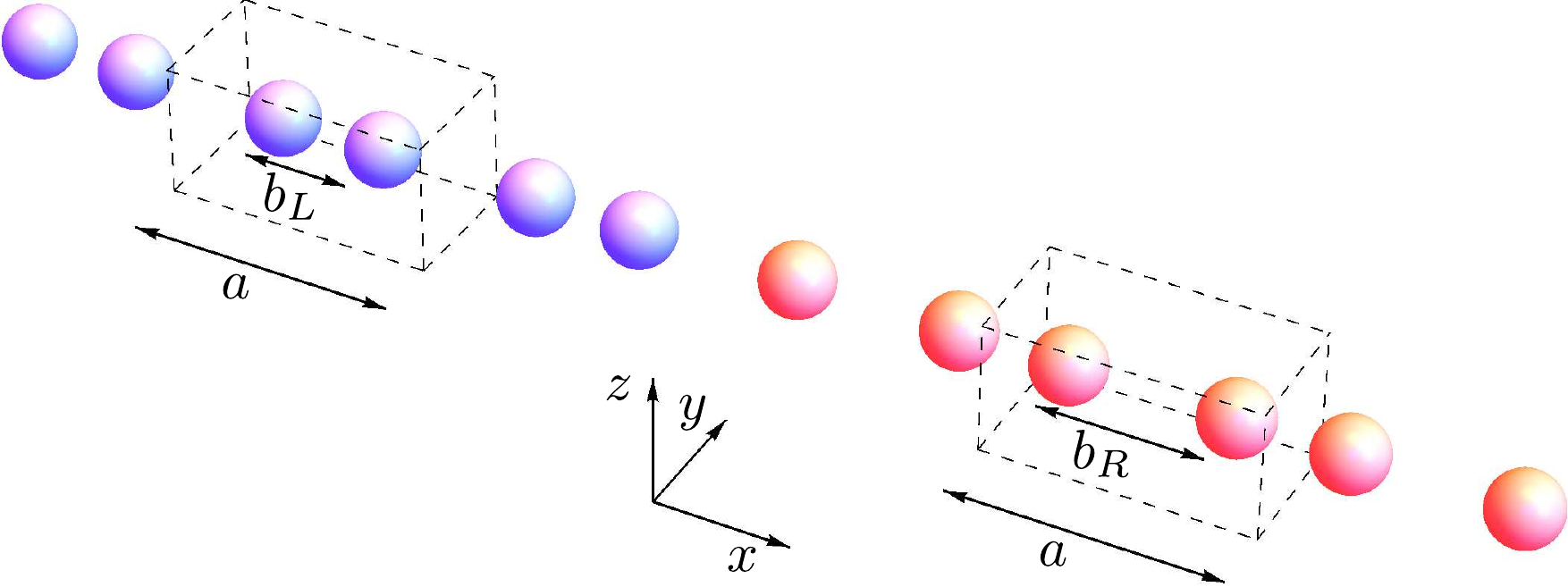}
    \caption{\label{fig12}Illustration of the 1-D topological plasmonic lattices.
    The finite lattice composed of a left and a right part with different unit cells.
    The unit cells are indicated by the dashed boxes.
    The lattice constant is given by $a$.
    The distance between nanoparticle $A$ and nanoparticle $B$, in the left and the right part of the lattice, is given by $b_L=(1-\delta_L)a/2$ and $b_R=(1-\delta_R)a/2$, respectively.
    }
\end{figure}
\begin{figure*}
    \includegraphics[width=17.8cm]{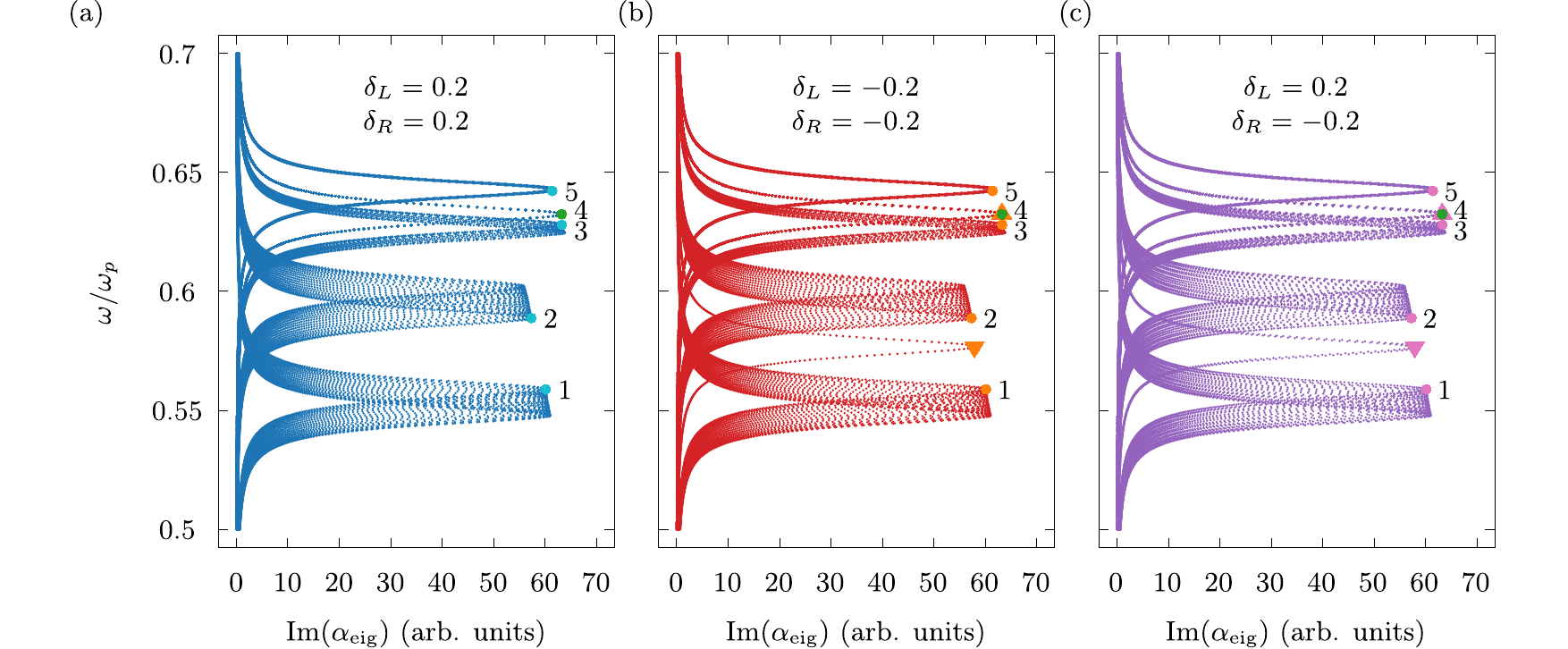}
    \caption{\label{fig13}Band structures of the 1-D finite bipartite lattices.
    (a) The topologically trivial system with both $\delta_L=0.2$ and $\delta_R=0.2$.
    (b) The topologically nontrivial system with both $\delta_L=-0.2$ and $\delta_R=-0.2$.
    (c) The system with $\delta_L=0.2$ and $\delta_R=-0.2$, where the left and the right part of the lattice are topologically trivial and nontrivial, respectively.
    The upper and the lower triangles indicate the high and low frequency topological edge states, respectively.
    The circles with band indexes indicate the bands correspond to the infinite lattices shown in Fig.~\ref{fig5}, in particular, the green circles indicates the quadrupolar flat bands.
    The results are calculated with $N=20$ and $\gamma=0.01\omega_p$.
    }
\end{figure*}
\begin{figure*}
    \includegraphics[width=17.8cm]{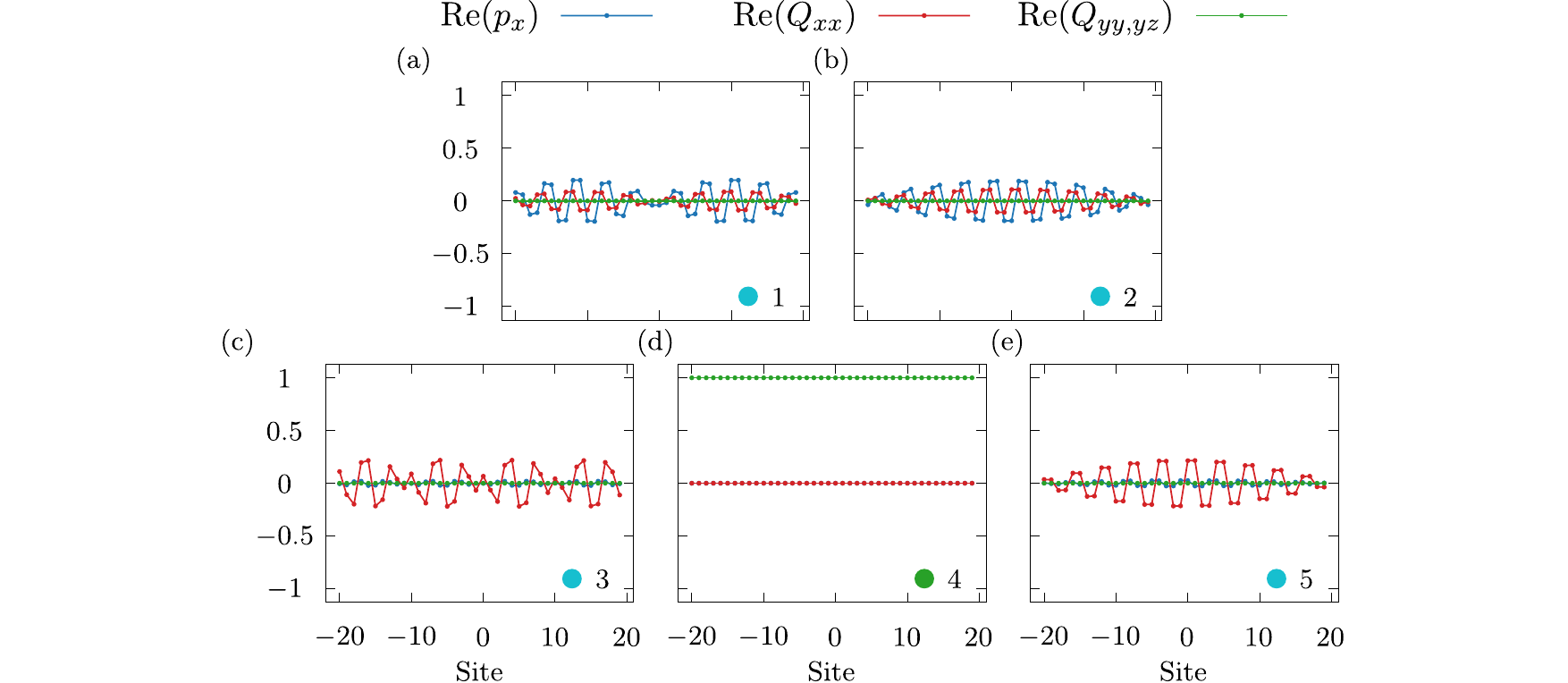}
    \caption{\label{fig14}The eigenmodes corresponding to the band structures in Fig.~\ref{fig13}(a).
    (a)--(c), (e) The bulk eigenmodes at the band edges of the trivial gaps.
    (d) The localized quadrupole modes.
    }
\end{figure*}
\begin{figure*}
    \includegraphics[width=17.8cm]{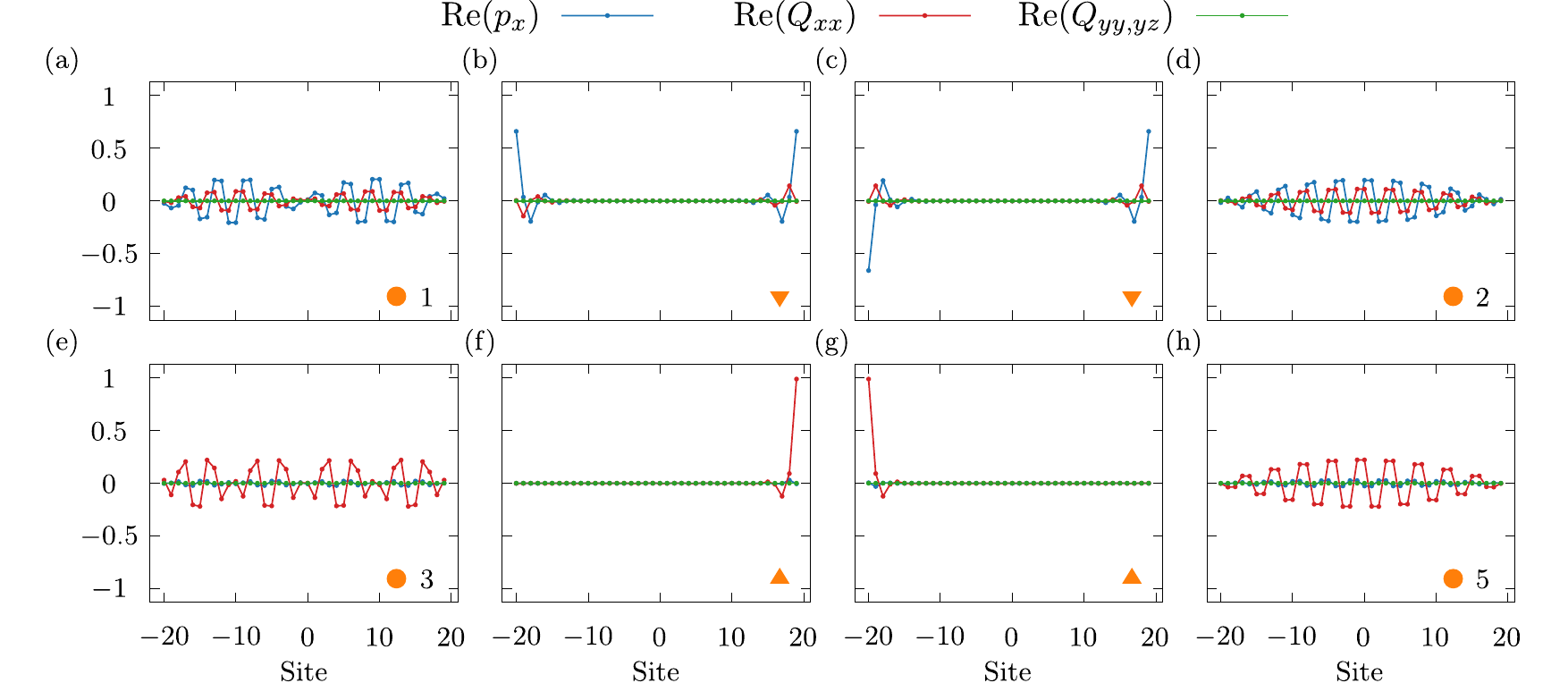}
    \caption{\label{fig15}The eigenmodes corresponding to the band structures in Fig.~\ref{fig13}(b).
    (a), (d), (e), (h) The bulk eigenmodes at the band edges of the non-trivial gaps.
    (b), (c) The coupled dipolar topological edge states.
    (f), (g) The uncoupled quadrupolar topological edge states.
    }
\end{figure*}
\begin{figure*}
    \includegraphics[width=17.8cm]{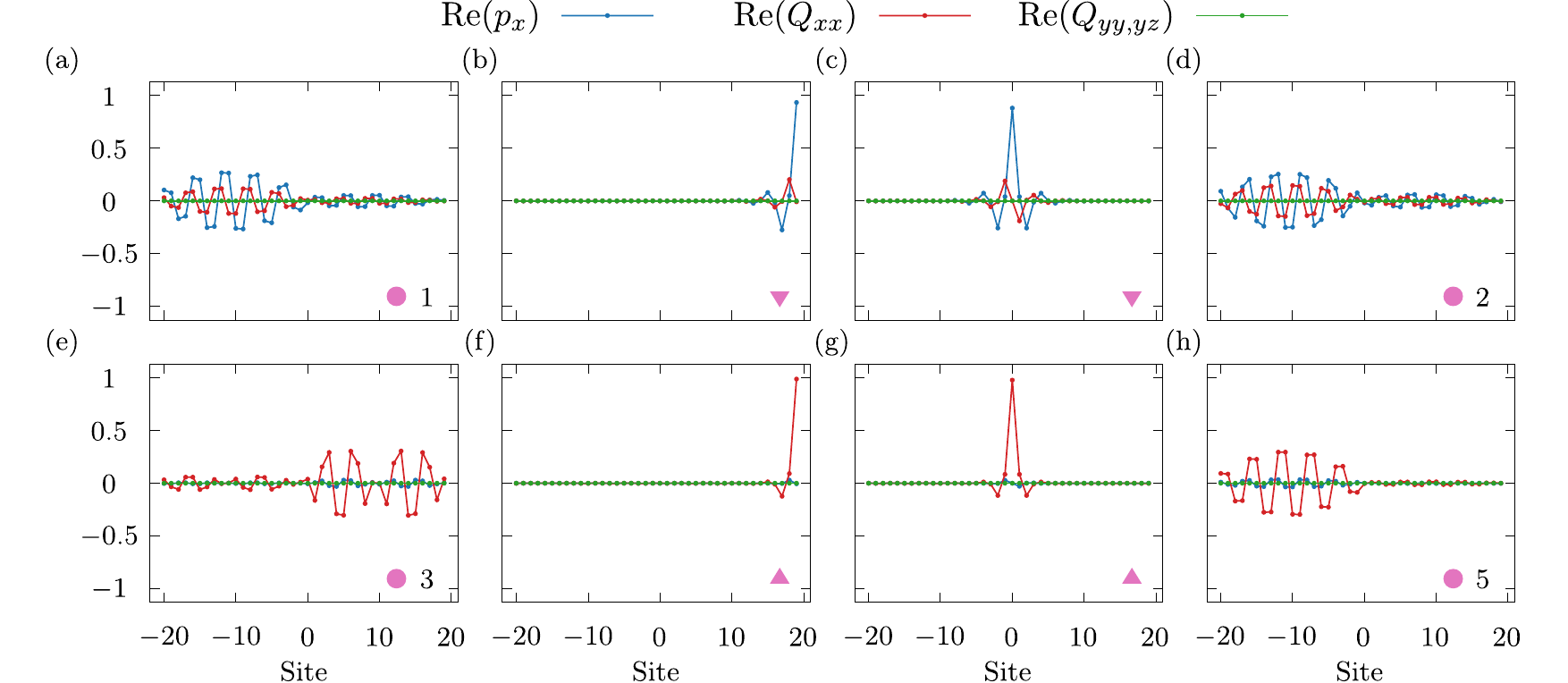}
    \caption{\label{fig16}The eigenmodes corresponding to the band structures in Fig.~\ref{fig13}(c).
    (a), (d), (e), (h) The bulk eigenmodes at the band edges of the non-trivial gaps.
    (b), (c) The dipolar topological edge states.
    (f), (g) The quadrupolar topological edge states.
    }
\end{figure*}
\begin{figure*}
    \includegraphics[width=17.8cm]{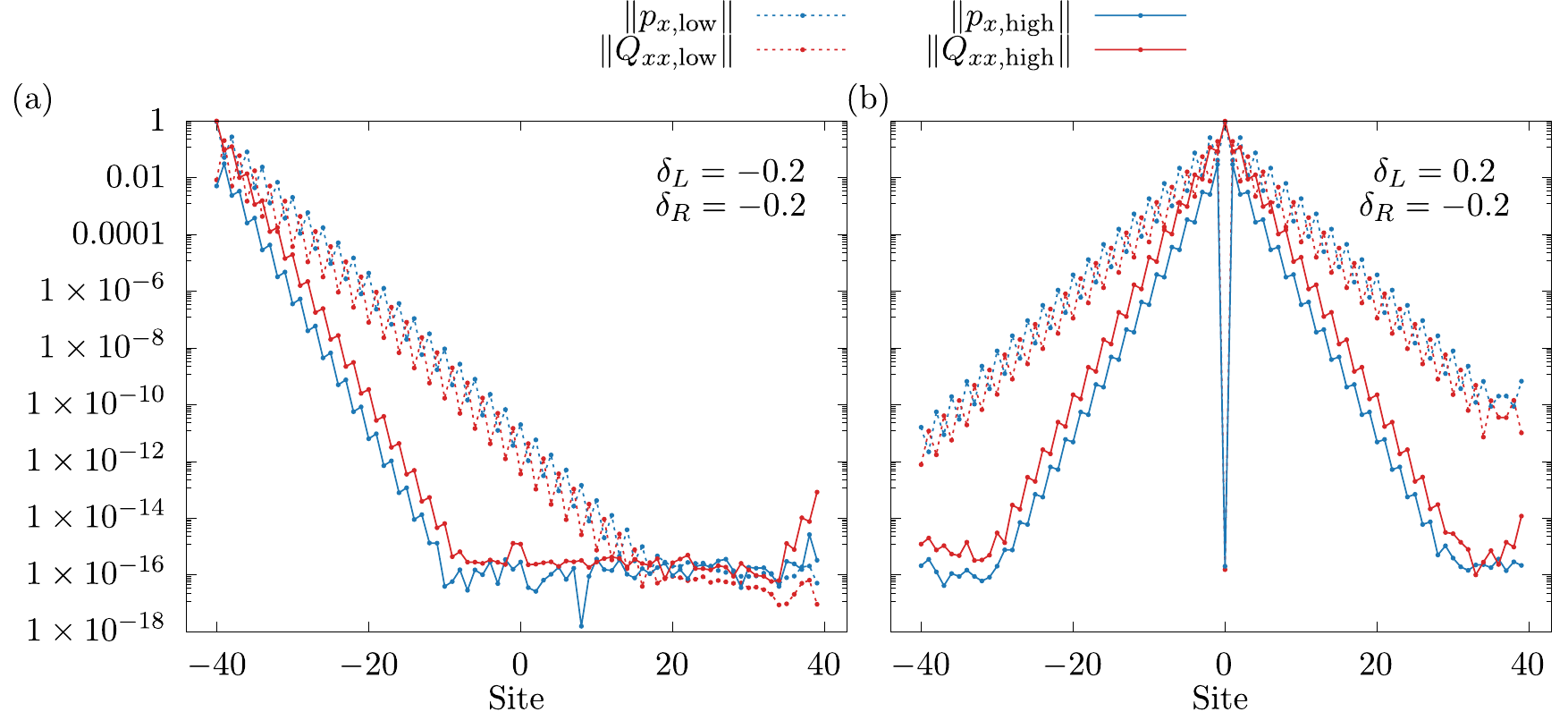}
    \caption{\label{fig17}The norm of topological edge states in $\log$ scale.
    The localization length $\xi$ is proportional to $1/\abs{m}$, where $m$ is the slope of the linear part of the envelopes.
    The high frequency quadrupolar topological edge states have a shorter localization length than the low frequency dipolar ones.
    The results are calculated with $N=40$ and $\gamma=0.01\omega_p$.
    }
\end{figure*}
To demonstrate the topological edge states, we consider the finite bipartite lattice model.
The finite lattice is composed of left and right parts with different unit cells.
The system is depicted in Fig.~\ref{fig12}.
We assume there is even $N$ unit cells and they are indexed by the integer $n=-N/2,\dots,0,\dots,N/2-1$.
Therefore $n<0$ corresponds to the left part and $n\geq0$ corresponds to the right part of the lattice.
The displacement from the nanoparticle $A$ to nanoparticle $B$ in the $n$th unit cell is given by $\bm{b}_n=b_n\hat{\bm{x}}$, where
\begin{equation}
    b_n=
    \begin{cases}
        b_L&\text{if $n<0$},\\
        b_R&\text{if $n\geq0$},
    \end{cases}\label{eqvi1}
\end{equation}
with
\begin{subequations}
    \label{eqvi2}
    \begin{eqnarray}
        b_L&=&\frac{a}{2}(1-\delta_L),\label{eqvi2a}
        \\
        b_R&=&\frac{a}{2}(1-\delta_R).\label{eqvi2b}
    \end{eqnarray}
\end{subequations}
In the $n$th unit cell, the state vector is $\bm{X}_n=(p_{A,x},Q_{A,xx},p_{B,x},Q_{B,xx})^T$ and the polarizability is $\bm{A}_n=\diag(\alpha_A^p,\alpha_A^Q,\alpha_B^p,\alpha_B^Q)$.
The interaction matrix can be constructed similar to that in Eq.~(\ref{eqv2}).
Generally, we have
\begin{equation}
    \bm{\Gamma}_{ij}=
    \begin{cases}
        \bm{0}&\text{if $i=j$},\\
        \bm{\Gamma}&\text{if $i\neq j$},
    \end{cases}\label{eqvi3}
\end{equation}
and after taking the nearest neighbor approximation, only the terms next to the diagonal remain.
Then for the finite lattices, we have the eigenvalue problem
\begin{equation}
    \bm{M}(\omega)\bm{X}_i=\lambda_i(\omega)\bm{X}_i,\label{eqvi4}
\end{equation}
where
\begin{equation}
    \bm{M}(\omega):=\bm{A}(\omega)-\bm{\Gamma},\label{eqvi5}
\end{equation}
is a $4N\times4N$ matrix.

We consider finite lattices with $N=20$.
First, we consider a system with $\delta_L=0.2$ and $\delta_R=0.2$.
This topologically trivial system is the finite case of the bipartite model discussed in Sec.~\ref{secv} with $\delta=0.2$.
The band structure for this finite system is shown in Fig.~\ref{fig13}(a) and their eigenmodes are shown in Fig.~\ref{fig14}.
We observe that there are four sets of bands and a quadrupolar flat band, which correspond to the bands in Fig.~\ref{fig5}(b), and their spectral positions are in good agreement.
The eigenmodes carry similar features from the infinite lattice, where the dipole moments and the quadrupole moments are always in different symmetry.
The dipole moments in a unit cell are in-phase in Band 1 and Band 3, while they are anti-phase in Band 2 and Band 5.
On the other hand, the quadrupole moments in a unit cell are in-phase in Band 2 and Band 5, and are anti-phase in Band 1 and Band 3.
In addition, the boundary conditions of the finite lattice lead to quantization of wavelengths\cite{weber2004}, which are different from the infinite cases in Fig.~\ref{fig6} and \ref{fig7}.

Next, we consider a topologically non-trivial system with $\delta_L=-0.2$ and $\delta_R=-0.2$.
Again, this is the finite case of the bipartite model discussed in Sec.~\ref{secv} with $\delta=-0.2$.
The band structure for this finite system is shown in Fig.~\ref{fig13}(b) and their eigenmodes are shown in Fig.~\ref{fig15}.
Similar to the trivial system, there are four sets of bands and a quadrupolar flat band, which correspond to the bands in Fig.~\ref{fig5}(b).
However, in contrast to the trivial case, there exist topological edge states at the non-trivial band gaps which we have identified in Sec.~\ref{secv}.
The bulk eigenmodes at the band edge of the non-trivial gap are shown in Fig.~\ref{fig15}(a), (d), (e), and (h).
The topological edge states within the non-trivial gaps are shown in Fig.~\ref{fig15}(b), (c), (f), and (g).
The topological edge states are degenerated and they localized at the end of the lattice exponentially.
We found that, the topological edge states have both dipolar and quadrupolar nature.
The topological edge states carry the same characteristics as the bulk eigenmodes, where the dipole moments and the quadrupole moments are spatially localized at different sublattices.
This is inherited from the $\pi/2$ phase difference between the dipole moments and the quadrupole moments in infinite lattices as discussed in Sec.~\ref{seciv}.

The dipolar topological edge states in Fig.~\ref{fig15}(b) and (c) are odd and even superpositions of states localized exponentially on the left and right edge.
This is the result of the exponentially small overlap between the left and the right edge states, which induces a small energy splitting, where the spectral positions of the topological edges states are almost at the resonant frequency of a single nanoparticle $\omega_0^p=\omega_p/\sqrt{3}$.
On the other hand, the quadrupolar topological edge states shown in Fig.~\ref{fig15}(f), (g) are uncoupled.
This suggests that the high frequency quadrupolar topological edge states have shorter localization lengths and hence different mode volume, when comparing to the low frequency dipolar ones.

We also consider another configuration with $\delta_L=0.2$ and $\delta_R=-0.2$, where the left part of the lattice is topologically trivial and the right part is non-trivial.
The band structure is shown in Fig.~\ref{fig13}(c) and their eigenmodes are shown in Fig.~\ref{fig16}.
Again there are topological edge states exist at the non-trivial band gaps.
The bulk eigenmodes at the band edge of the non-trivial gap are shown in Fig.~\ref{fig16}(a), (d), (e), and (h).
The topological edge states within the non-trivial gaps are shown in Fig.~\ref{fig16}(b), (c), (f), and (g), with the dipolar ones given in Fig.~\ref{fig16}(b), (c), and the quadripolar ones given in Fig.~\ref{fig16}(f), (g).
Both topological edge states are degenerated with one solution localized at the center and another localized at the right end, which are the positions where mismatch between the band topologies occurred.

In both Fig.~\ref{fig13}(b) and (c), the low frequency topological edge states appear at the dipolar resonant frequency of the nanoparticles $\omega_0^p=\omega_p/\sqrt{3}$, while the high frequency ones appear at the quadrupolar resonant frequency of the nanoparticles $\omega_0^Q=\omega_p\sqrt{2/5}$.
This is a consequence of the chiral symmetry, where the spectral positions of the topological edge states are at the zero-energy states\cite{asboth2016}.
As long as only nearest neighbor interactions are included in the calculations, chiral symmetry is present\cite{pocock2018}.
As a result, the quadrupolar topological edge states always coexist at the same energy with the quadrupolar flat band in 1-D lattices.

To verify the localization lengths of the topological edge states, we consider finite lattices with $N=40$.
In particular, for the system with $\delta_L=-0.2$ and $\delta_R=-0.2$, the topological edge states localized at the left end are chosen to study, while for the system with $\delta_L=0.2$ and $\delta_R=-0.2$, the ones that localized at the center are chosen.
The norm of these topological edge states, given by $\norm{p_x}=\sqrt{p_x^*p_x}$ and $\norm{Q_{xx}}=\sqrt{Q_{xx}^*Q_{xx}}$, are plotted in $\log$ scale in Fig.~\ref{fig17}.
In SSH model, the localization length $\xi$ of the topological edge state depends on the strength of the inter- and intra-cell interactions and it can be obtained from $\xi=1/\abs{m}$, where $m$ is the slope of the envelope\cite{asboth2016,obana2019}.
In Fig.~\ref{fig17}(a), by fitting the linear part of the envelope, the localization length for the low frequency dipolar edge state is found to be $\xi=1.628$, while that for the high frequency quadrupolar edge state is $\xi=0.915$, which is only $56.199\%$ of the dipolar one.
Similarly, in Fig.~\ref{fig17}(b), the localization lengths for both $n<0$ and $n>0$ are approximately the same, with $\xi\approx1.618$ for the low frequency dipolar edge state and $\xi\approx0.898$ for the high frequency quadrupolar edge state, which is only $55.509\%$ of the dipolar one.
The dipole--quadrupole interactions lead to a high frequency quadrupolar edge states with smaller mode volume when comparing to the low frequency dipolar one.
Hence, topological edge states arise from multipolar interactions provides an alternative way to confine light with small mode volume while at the same time are topologically protected.

\section{\label{secvii}Conclusion}
We studied the topological photonic states in 1-D lattices analogue to the SSH model with coupled dipole--quadrupole method.
Our work extended previous works on plasmonic nanoparticles in 1-D lattices by including all the dipole--dipole, quadrupole--quadrupole, and dipole--quadrupole interactions.
Our results reveal the contribution of quadrupole moments to the near-field interactions and the band topology.
The topological edge states are found to have both dipolar and quadripolar nature.
Due to the $\pi/2$ phase difference between the dipole moments and the quadrupole moments, the quadrupolar edge states are not only orthogonal to the dipolar edge states, but also spatially localized at different sublattices.
The quadrupolar topological edge states, which coexist at the same energy with the quadrupolar flat band have shorter localization length and hence smaller mode volume than the conventional dipolar edge states.
The findings deepen our understanding in topological systems that involve higher-order multipoles, or in analogy to the wave functions in quantum systems with higher-orbital angular momentum, and may be useful in designing topological systems for confining light robustly and enhancing light-matter interactions.

\begin{acknowledgments}
    This research was supported by the Chinese University of Hong Kong through Area of Excellence (AoE/P-02/12) and Innovative Technology Funds (ITS/133/19 and UIM/397).
\end{acknowledgments}

\begin{thebibliography}{62}%
    \makeatletter
    \providecommand \@ifxundefined [1]{%
     \@ifx{#1\undefined}
    }%
    \providecommand \@ifnum [1]{%
     \ifnum #1\expandafter \@firstoftwo
     \else \expandafter \@secondoftwo
     \fi
    }%
    \providecommand \@ifx [1]{%
     \ifx #1\expandafter \@firstoftwo
     \else \expandafter \@secondoftwo
     \fi
    }%
    \providecommand \natexlab [1]{#1}%
    \providecommand \enquote  [1]{``#1''}%
    \providecommand \bibnamefont  [1]{#1}%
    \providecommand \bibfnamefont [1]{#1}%
    \providecommand \citenamefont [1]{#1}%
    \providecommand \href@noop [0]{\@secondoftwo}%
    \providecommand \href [0]{\begingroup \@sanitize@url \@href}%
    \providecommand \@href[1]{\@@startlink{#1}\@@href}%
    \providecommand \@@href[1]{\endgroup#1\@@endlink}%
    \providecommand \@sanitize@url [0]{\catcode `\\12\catcode `\$12\catcode
      `\&12\catcode `\#12\catcode `\^12\catcode `\_12\catcode `\%12\relax}%
    \providecommand \@@startlink[1]{}%
    \providecommand \@@endlink[0]{}%
    \providecommand \url  [0]{\begingroup\@sanitize@url \@url }%
    \providecommand \@url [1]{\endgroup\@href {#1}{\urlprefix }}%
    \providecommand \urlprefix  [0]{URL }%
    \providecommand \Eprint [0]{\href }%
    \providecommand \doibase [0]{https://doi.org/}%
    \providecommand \selectlanguage [0]{\@gobble}%
    \providecommand \bibinfo  [0]{\@secondoftwo}%
    \providecommand \bibfield  [0]{\@secondoftwo}%
    \providecommand \translation [1]{[#1]}%
    \providecommand \BibitemOpen [0]{}%
    \providecommand \bibitemStop [0]{}%
    \providecommand \bibitemNoStop [0]{.\EOS\space}%
    \providecommand \EOS [0]{\spacefactor3000\relax}%
    \providecommand \BibitemShut  [1]{\csname bibitem#1\endcsname}%
    \let\auto@bib@innerbib\@empty
    \bibitem [{\citenamefont {Hasan}\ and\ \citenamefont {Kane}(2010)}]{hasan2010}%
      \BibitemOpen
      \bibfield  {author} {\bibinfo {author} {\bibfnamefont {M.~Z.}\ \bibnamefont
      {Hasan}}\ and\ \bibinfo {author} {\bibfnamefont {C.~L.}\ \bibnamefont
      {Kane}},\ }\bibfield  {title} {\bibinfo {title} {Colloquium: Topological
      insulators},\ }\href {https://doi.org/10.1103/RevModPhys.82.3045} {\bibfield
      {journal} {\bibinfo  {journal} {Rev. Mod. Phys.}\ }\textbf {\bibinfo {volume}
      {82}},\ \bibinfo {pages} {3045} (\bibinfo {year} {2010})}\BibitemShut
      {NoStop}%
    \bibitem [{\citenamefont {Wang}\ \emph {et~al.}(2019)\citenamefont {Wang},
      \citenamefont {Guo},\ and\ \citenamefont {Jiang}}]{wang2019}%
      \BibitemOpen
      \bibfield  {author} {\bibinfo {author} {\bibfnamefont {H.-X.}\ \bibnamefont
      {Wang}}, \bibinfo {author} {\bibfnamefont {G.-Y.}\ \bibnamefont {Guo}},\ and\
      \bibinfo {author} {\bibfnamefont {J.-H.}\ \bibnamefont {Jiang}},\ }\bibfield
      {title} {\bibinfo {title} {Band topology in classical waves: Wilson-loop
      approach to topological numbers and fragile topology},\ }\href
      {https://doi.org/10.1088/1367-2630/ab3f71} {\bibfield  {journal} {\bibinfo
      {journal} {New J. Phys.}\ }\textbf {\bibinfo {volume} {21}},\ \bibinfo
      {pages} {093029} (\bibinfo {year} {2019})}\BibitemShut {NoStop}%
    \bibitem [{\citenamefont {Ozawa}\ \emph {et~al.}(2019)\citenamefont {Ozawa},
      \citenamefont {Price}, \citenamefont {Amo}, \citenamefont {Goldman},
      \citenamefont {Hafezi}, \citenamefont {Lu}, \citenamefont {Rechtsman},
      \citenamefont {Schuster}, \citenamefont {Simon}, \citenamefont {Zilberberg},\
      and\ \citenamefont {Carusotto}}]{ozawa2019}%
      \BibitemOpen
      \bibfield  {author} {\bibinfo {author} {\bibfnamefont {T.}~\bibnamefont
      {Ozawa}}, \bibinfo {author} {\bibfnamefont {H.~M.}\ \bibnamefont {Price}},
      \bibinfo {author} {\bibfnamefont {A.}~\bibnamefont {Amo}}, \bibinfo {author}
      {\bibfnamefont {N.}~\bibnamefont {Goldman}}, \bibinfo {author} {\bibfnamefont
      {M.}~\bibnamefont {Hafezi}}, \bibinfo {author} {\bibfnamefont
      {L.}~\bibnamefont {Lu}}, \bibinfo {author} {\bibfnamefont {M.~C.}\
      \bibnamefont {Rechtsman}}, \bibinfo {author} {\bibfnamefont {D.}~\bibnamefont
      {Schuster}}, \bibinfo {author} {\bibfnamefont {J.}~\bibnamefont {Simon}},
      \bibinfo {author} {\bibfnamefont {O.}~\bibnamefont {Zilberberg}},\ and\
      \bibinfo {author} {\bibfnamefont {I.}~\bibnamefont {Carusotto}},\ }\bibfield
      {title} {\bibinfo {title} {Topological photonics},\ }\href
      {https://doi.org/10.1103/RevModPhys.91.015006} {\bibfield  {journal}
      {\bibinfo  {journal} {Rev. Mod. Phys.}\ }\textbf {\bibinfo {volume} {91}},\
      \bibinfo {pages} {015006} (\bibinfo {year} {2019})}\BibitemShut {NoStop}%
    \bibitem [{\citenamefont {Haldane}\ and\ \citenamefont
      {Raghu}(2008)}]{haldane2008}%
      \BibitemOpen
      \bibfield  {author} {\bibinfo {author} {\bibfnamefont {F.~D.~M.}\
      \bibnamefont {Haldane}}\ and\ \bibinfo {author} {\bibfnamefont
      {S.}~\bibnamefont {Raghu}},\ }\bibfield  {title} {\bibinfo {title} {Possible
      realization of directional optical waveguides in photonic crystals with
      broken time-reversal symmetry},\ }\href
      {https://doi.org/10.1103/PhysRevLett.100.013904} {\bibfield  {journal}
      {\bibinfo  {journal} {Phys. Rev. Lett.}\ }\textbf {\bibinfo {volume} {100}},\
      \bibinfo {pages} {013904} (\bibinfo {year} {2008})}\BibitemShut {NoStop}%
    \bibitem [{\citenamefont {Wang}\ \emph {et~al.}(2008)\citenamefont {Wang},
      \citenamefont {Chong}, \citenamefont {Joannopoulos},\ and\ \citenamefont
      {Solja\ifmmode \check{c}\else \v{c}\fi{}i\ifmmode~\acute{c}\else
      \'{c}\fi{}}}]{wang2008}%
      \BibitemOpen
      \bibfield  {author} {\bibinfo {author} {\bibfnamefont {Z.}~\bibnamefont
      {Wang}}, \bibinfo {author} {\bibfnamefont {Y.~D.}\ \bibnamefont {Chong}},
      \bibinfo {author} {\bibfnamefont {J.~D.}\ \bibnamefont {Joannopoulos}},\ and\
      \bibinfo {author} {\bibfnamefont {M.}~\bibnamefont {Solja\ifmmode
      \check{c}\else \v{c}\fi{}i\ifmmode~\acute{c}\else \'{c}\fi{}}},\ }\bibfield
      {title} {\bibinfo {title} {Reflection-free one-way edge modes in a
      gyromagnetic photonic crystal},\ }\href
      {https://doi.org/10.1103/PhysRevLett.100.013905} {\bibfield  {journal}
      {\bibinfo  {journal} {Phys. Rev. Lett.}\ }\textbf {\bibinfo {volume} {100}},\
      \bibinfo {pages} {013905} (\bibinfo {year} {2008})}\BibitemShut {NoStop}%
    \bibitem [{\citenamefont {Raghu}\ and\ \citenamefont
      {Haldane}(2008)}]{raghu2008}%
      \BibitemOpen
      \bibfield  {author} {\bibinfo {author} {\bibfnamefont {S.}~\bibnamefont
      {Raghu}}\ and\ \bibinfo {author} {\bibfnamefont {F.~D.~M.}\ \bibnamefont
      {Haldane}},\ }\bibfield  {title} {\bibinfo {title} {Analogs of
      quantum-hall-effect edge states in photonic crystals},\ }\href
      {https://doi.org/10.1103/PhysRevA.78.033834} {\bibfield  {journal} {\bibinfo
      {journal} {Phys. Rev. A}\ }\textbf {\bibinfo {volume} {78}},\ \bibinfo
      {pages} {033834} (\bibinfo {year} {2008})}\BibitemShut {NoStop}%
    \bibitem [{\citenamefont {Wang}\ \emph {et~al.}(2009)\citenamefont {Wang},
      \citenamefont {Chong}, \citenamefont {Joannopoulos},\ and\ \citenamefont
      {Solja{\v{c}}i{\'c}}}]{wang2009}%
      \BibitemOpen
      \bibfield  {author} {\bibinfo {author} {\bibfnamefont {Z.}~\bibnamefont
      {Wang}}, \bibinfo {author} {\bibfnamefont {Y.}~\bibnamefont {Chong}},
      \bibinfo {author} {\bibfnamefont {J.~D.}\ \bibnamefont {Joannopoulos}},\ and\
      \bibinfo {author} {\bibfnamefont {M.}~\bibnamefont {Solja{\v{c}}i{\'c}}},\
      }\bibfield  {title} {\bibinfo {title} {Observation of unidirectional
      backscattering-immune topological electromagnetic states},\ }\href@noop {}
      {\bibfield  {journal} {\bibinfo  {journal} {Nature}\ }\textbf {\bibinfo
      {volume} {461}},\ \bibinfo {pages} {772} (\bibinfo {year}
      {2009})}\BibitemShut {NoStop}%
    \bibitem [{\citenamefont {Su}\ \emph {et~al.}(1979)\citenamefont {Su},
      \citenamefont {Schrieffer},\ and\ \citenamefont {Heeger}}]{su1979}%
      \BibitemOpen
      \bibfield  {author} {\bibinfo {author} {\bibfnamefont {W.~P.}\ \bibnamefont
      {Su}}, \bibinfo {author} {\bibfnamefont {J.~R.}\ \bibnamefont {Schrieffer}},\
      and\ \bibinfo {author} {\bibfnamefont {A.~J.}\ \bibnamefont {Heeger}},\
      }\bibfield  {title} {\bibinfo {title} {Solitons in polyacetylene},\ }\href
      {https://doi.org/10.1103/PhysRevLett.42.1698} {\bibfield  {journal} {\bibinfo
       {journal} {Phys. Rev. Lett.}\ }\textbf {\bibinfo {volume} {42}},\ \bibinfo
      {pages} {1698} (\bibinfo {year} {1979})}\BibitemShut {NoStop}%
    \bibitem [{\citenamefont {Keil}\ \emph {et~al.}(2013)\citenamefont {Keil},
      \citenamefont {Zeuner}, \citenamefont {Dreisow}, \citenamefont {Heinrich},
      \citenamefont {T{\"u}nnermann}, \citenamefont {Nolte},\ and\ \citenamefont
      {Szameit}}]{keil2013}%
      \BibitemOpen
      \bibfield  {author} {\bibinfo {author} {\bibfnamefont {R.}~\bibnamefont
      {Keil}}, \bibinfo {author} {\bibfnamefont {J.~M.}\ \bibnamefont {Zeuner}},
      \bibinfo {author} {\bibfnamefont {F.}~\bibnamefont {Dreisow}}, \bibinfo
      {author} {\bibfnamefont {M.}~\bibnamefont {Heinrich}}, \bibinfo {author}
      {\bibfnamefont {A.}~\bibnamefont {T{\"u}nnermann}}, \bibinfo {author}
      {\bibfnamefont {S.}~\bibnamefont {Nolte}},\ and\ \bibinfo {author}
      {\bibfnamefont {A.}~\bibnamefont {Szameit}},\ }\bibfield  {title} {\bibinfo
      {title} {The random mass dirac model and long-range correlations on an
      integrated optical platform},\ }\href@noop {} {\bibfield  {journal} {\bibinfo
       {journal} {Nat. Commun.}\ }\textbf {\bibinfo {volume} {4}},\ \bibinfo
      {pages} {1368} (\bibinfo {year} {2013})}\BibitemShut {NoStop}%
    \bibitem [{\citenamefont {Xiao}\ \emph {et~al.}(2014)\citenamefont {Xiao},
      \citenamefont {Zhang},\ and\ \citenamefont {Chan}}]{xiao2014}%
      \BibitemOpen
      \bibfield  {author} {\bibinfo {author} {\bibfnamefont {M.}~\bibnamefont
      {Xiao}}, \bibinfo {author} {\bibfnamefont {Z.~Q.}\ \bibnamefont {Zhang}},\
      and\ \bibinfo {author} {\bibfnamefont {C.~T.}\ \bibnamefont {Chan}},\
      }\bibfield  {title} {\bibinfo {title} {Surface impedance and bulk band
      geometric phases in one-dimensional systems},\ }\href
      {https://doi.org/10.1103/PhysRevX.4.021017} {\bibfield  {journal} {\bibinfo
      {journal} {Phys. Rev. X}\ }\textbf {\bibinfo {volume} {4}},\ \bibinfo {pages}
      {021017} (\bibinfo {year} {2014})}\BibitemShut {NoStop}%
    \bibitem [{\citenamefont {Poddubny}\ \emph {et~al.}(2014)\citenamefont
      {Poddubny}, \citenamefont {Miroshnichenko}, \citenamefont {Slobozhanyuk},\
      and\ \citenamefont {Kivshar}}]{poddubny2014}%
      \BibitemOpen
      \bibfield  {author} {\bibinfo {author} {\bibfnamefont {A.}~\bibnamefont
      {Poddubny}}, \bibinfo {author} {\bibfnamefont {A.}~\bibnamefont
      {Miroshnichenko}}, \bibinfo {author} {\bibfnamefont {A.}~\bibnamefont
      {Slobozhanyuk}},\ and\ \bibinfo {author} {\bibfnamefont {Y.}~\bibnamefont
      {Kivshar}},\ }\bibfield  {title} {\bibinfo {title} {Topological majorana
      states in zigzag chains of plasmonic nanoparticles},\ }\href
      {https://doi.org/10.1021/ph4000949} {\bibfield  {journal} {\bibinfo
      {journal} {ACS Photonics}\ }\textbf {\bibinfo {volume} {1}},\ \bibinfo
      {pages} {101} (\bibinfo {year} {2014})},\ \Eprint
      {https://arxiv.org/abs/https://doi.org/10.1021/ph4000949}
      {https://doi.org/10.1021/ph4000949} \BibitemShut {NoStop}%
    \bibitem [{\citenamefont {Ling}\ \emph {et~al.}(2015)\citenamefont {Ling},
      \citenamefont {Xiao}, \citenamefont {Chan}, \citenamefont {Yu},\ and\
      \citenamefont {Fung}}]{ling2015}%
      \BibitemOpen
      \bibfield  {author} {\bibinfo {author} {\bibfnamefont {C.~W.}\ \bibnamefont
      {Ling}}, \bibinfo {author} {\bibfnamefont {M.}~\bibnamefont {Xiao}}, \bibinfo
      {author} {\bibfnamefont {C.~T.}\ \bibnamefont {Chan}}, \bibinfo {author}
      {\bibfnamefont {S.~F.}\ \bibnamefont {Yu}},\ and\ \bibinfo {author}
      {\bibfnamefont {K.~H.}\ \bibnamefont {Fung}},\ }\bibfield  {title} {\bibinfo
      {title} {Topological edge plasmon modes between diatomic chains of plasmonic
      nanoparticles},\ }\href {https://doi.org/10.1364/OE.23.002021} {\bibfield
      {journal} {\bibinfo  {journal} {Opt. Express}\ }\textbf {\bibinfo {volume}
      {23}},\ \bibinfo {pages} {2021} (\bibinfo {year} {2015})}\BibitemShut
      {NoStop}%
    \bibitem [{\citenamefont {Sinev}\ \emph {et~al.}(2015)\citenamefont {Sinev},
      \citenamefont {Mukhin}, \citenamefont {Slobozhanyuk}, \citenamefont
      {Poddubny}, \citenamefont {Miroshnichenko}, \citenamefont {Samusev},\ and\
      \citenamefont {Kivshar}}]{sinev2015}%
      \BibitemOpen
      \bibfield  {author} {\bibinfo {author} {\bibfnamefont {I.~S.}\ \bibnamefont
      {Sinev}}, \bibinfo {author} {\bibfnamefont {I.~S.}\ \bibnamefont {Mukhin}},
      \bibinfo {author} {\bibfnamefont {A.~P.}\ \bibnamefont {Slobozhanyuk}},
      \bibinfo {author} {\bibfnamefont {A.~N.}\ \bibnamefont {Poddubny}}, \bibinfo
      {author} {\bibfnamefont {A.~E.}\ \bibnamefont {Miroshnichenko}}, \bibinfo
      {author} {\bibfnamefont {A.~K.}\ \bibnamefont {Samusev}},\ and\ \bibinfo
      {author} {\bibfnamefont {Y.~S.}\ \bibnamefont {Kivshar}},\ }\bibfield
      {title} {\bibinfo {title} {Mapping plasmonic topological states at the
      nanoscale},\ }\href {https://doi.org/10.1039/C5NR00231A} {\bibfield
      {journal} {\bibinfo  {journal} {Nanoscale}\ }\textbf {\bibinfo {volume}
      {7}},\ \bibinfo {pages} {11904} (\bibinfo {year} {2015})}\BibitemShut
      {NoStop}%
    \bibitem [{\citenamefont {Downing}\ and\ \citenamefont
      {Weick}(2017)}]{downing2017}%
      \BibitemOpen
      \bibfield  {author} {\bibinfo {author} {\bibfnamefont {C.~A.}\ \bibnamefont
      {Downing}}\ and\ \bibinfo {author} {\bibfnamefont {G.}~\bibnamefont
      {Weick}},\ }\bibfield  {title} {\bibinfo {title} {Topological collective
      plasmons in bipartite chains of metallic nanoparticles},\ }\href
      {https://doi.org/10.1103/PhysRevB.95.125426} {\bibfield  {journal} {\bibinfo
      {journal} {Phys. Rev. B}\ }\textbf {\bibinfo {volume} {95}},\ \bibinfo
      {pages} {125426} (\bibinfo {year} {2017})}\BibitemShut {NoStop}%
    \bibitem [{\citenamefont {Zhang}\ \emph {et~al.}(2018)\citenamefont {Zhang},
      \citenamefont {Wu}, \citenamefont {Kumar}, \citenamefont {Si},\ and\
      \citenamefont {Fung}}]{zhang2018}%
      \BibitemOpen
      \bibfield  {author} {\bibinfo {author} {\bibfnamefont {Y.-L.}\ \bibnamefont
      {Zhang}}, \bibinfo {author} {\bibfnamefont {R.~P.~H.}\ \bibnamefont {Wu}},
      \bibinfo {author} {\bibfnamefont {A.}~\bibnamefont {Kumar}}, \bibinfo
      {author} {\bibfnamefont {T.}~\bibnamefont {Si}},\ and\ \bibinfo {author}
      {\bibfnamefont {K.~H.}\ \bibnamefont {Fung}},\ }\bibfield  {title} {\bibinfo
      {title} {Nonsymmorphic symmetry-protected topological modes in plasmonic
      nanoribbon lattices},\ }\href {https://doi.org/10.1103/PhysRevB.97.144203}
      {\bibfield  {journal} {\bibinfo  {journal} {Phys. Rev. B}\ }\textbf {\bibinfo
      {volume} {97}},\ \bibinfo {pages} {144203} (\bibinfo {year}
      {2018})}\BibitemShut {NoStop}%
    \bibitem [{\citenamefont {Pocock}\ \emph {et~al.}(2018)\citenamefont {Pocock},
      \citenamefont {Xiao}, \citenamefont {Huidobro},\ and\ \citenamefont
      {Giannini}}]{pocock2018}%
      \BibitemOpen
      \bibfield  {author} {\bibinfo {author} {\bibfnamefont {S.~R.}\ \bibnamefont
      {Pocock}}, \bibinfo {author} {\bibfnamefont {X.}~\bibnamefont {Xiao}},
      \bibinfo {author} {\bibfnamefont {P.~A.}\ \bibnamefont {Huidobro}},\ and\
      \bibinfo {author} {\bibfnamefont {V.}~\bibnamefont {Giannini}},\ }\bibfield
      {title} {\bibinfo {title} {Topological plasmonic chain with retardation and
      radiative effects},\ }\href {https://doi.org/10.1021/acsphotonics.8b00117}
      {\bibfield  {journal} {\bibinfo  {journal} {ACS Photonics}\ }\textbf
      {\bibinfo {volume} {5}},\ \bibinfo {pages} {2271} (\bibinfo {year} {2018})},\
      \Eprint {https://arxiv.org/abs/https://doi.org/10.1021/acsphotonics.8b00117}
      {https://doi.org/10.1021/acsphotonics.8b00117} \BibitemShut {NoStop}%
    \bibitem [{\citenamefont {Downing}\ and\ \citenamefont
      {Weick}(2018)}]{downing2018}%
      \BibitemOpen
      \bibfield  {author} {\bibinfo {author} {\bibfnamefont {C.~A.}\ \bibnamefont
      {Downing}}\ and\ \bibinfo {author} {\bibfnamefont {G.}~\bibnamefont
      {Weick}},\ }\bibfield  {title} {\bibinfo {title} {Topological plasmons in
      dimerized chains of nanoparticles: robustness against long-range quasistatic
      interactions and retardation effects},\ }\href@noop {} {\bibfield  {journal}
      {\bibinfo  {journal} {Eur. Phys. J. B}\ }\textbf {\bibinfo {volume} {91}},\
      \bibinfo {pages} {253} (\bibinfo {year} {2018})}\BibitemShut {NoStop}%
    \bibitem [{\citenamefont {Slobozhanyuk}\ \emph {et~al.}(2015)\citenamefont
      {Slobozhanyuk}, \citenamefont {Poddubny}, \citenamefont {Miroshnichenko},
      \citenamefont {Belov},\ and\ \citenamefont {Kivshar}}]{slobozhanyuk2015}%
      \BibitemOpen
      \bibfield  {author} {\bibinfo {author} {\bibfnamefont {A.~P.}\ \bibnamefont
      {Slobozhanyuk}}, \bibinfo {author} {\bibfnamefont {A.~N.}\ \bibnamefont
      {Poddubny}}, \bibinfo {author} {\bibfnamefont {A.~E.}\ \bibnamefont
      {Miroshnichenko}}, \bibinfo {author} {\bibfnamefont {P.~A.}\ \bibnamefont
      {Belov}},\ and\ \bibinfo {author} {\bibfnamefont {Y.~S.}\ \bibnamefont
      {Kivshar}},\ }\bibfield  {title} {\bibinfo {title} {Subwavelength topological
      edge states in optically resonant dielectric structures},\ }\href
      {https://doi.org/10.1103/PhysRevLett.114.123901} {\bibfield  {journal}
      {\bibinfo  {journal} {Phys. Rev. Lett.}\ }\textbf {\bibinfo {volume} {114}},\
      \bibinfo {pages} {123901} (\bibinfo {year} {2015})}\BibitemShut {NoStop}%
    \bibitem [{\citenamefont {Slobozhanyuk}\ \emph {et~al.}(2016)\citenamefont
      {Slobozhanyuk}, \citenamefont {Poddubny}, \citenamefont {Sinev},
      \citenamefont {Samusev}, \citenamefont {Yu}, \citenamefont {Kuznetsov},
      \citenamefont {Miroshnichenko},\ and\ \citenamefont
      {Kivshar}}]{slobozhanyuk2016}%
      \BibitemOpen
      \bibfield  {author} {\bibinfo {author} {\bibfnamefont {A.~P.}\ \bibnamefont
      {Slobozhanyuk}}, \bibinfo {author} {\bibfnamefont {A.~N.}\ \bibnamefont
      {Poddubny}}, \bibinfo {author} {\bibfnamefont {I.~S.}\ \bibnamefont {Sinev}},
      \bibinfo {author} {\bibfnamefont {A.~K.}\ \bibnamefont {Samusev}}, \bibinfo
      {author} {\bibfnamefont {Y.~F.}\ \bibnamefont {Yu}}, \bibinfo {author}
      {\bibfnamefont {A.~I.}\ \bibnamefont {Kuznetsov}}, \bibinfo {author}
      {\bibfnamefont {A.~E.}\ \bibnamefont {Miroshnichenko}},\ and\ \bibinfo
      {author} {\bibfnamefont {Y.~S.}\ \bibnamefont {Kivshar}},\ }\bibfield
      {title} {\bibinfo {title} {Enhanced photonic spin hall effect with
      subwavelength topological edge states},\ }\href
      {https://doi.org/https://doi.org/10.1002/lpor.201600042} {\bibfield
      {journal} {\bibinfo  {journal} {Laser \& Photonics Reviews}\ }\textbf
      {\bibinfo {volume} {10}},\ \bibinfo {pages} {656} (\bibinfo {year} {2016})},\
      \Eprint
      {https://arxiv.org/abs/https://onlinelibrary.wiley.com/doi/pdf/10.1002/lpor.201600042}
      {https://onlinelibrary.wiley.com/doi/pdf/10.1002/lpor.201600042} \BibitemShut
      {NoStop}%
    \bibitem [{\citenamefont {Kruk}\ \emph {et~al.}(2017)\citenamefont {Kruk},
      \citenamefont {Slobozhanyuk}, \citenamefont {Denkova}, \citenamefont
      {Poddubny}, \citenamefont {Kravchenko}, \citenamefont {Miroshnichenko},
      \citenamefont {Neshev},\ and\ \citenamefont {Kivshar}}]{kruk2017}%
      \BibitemOpen
      \bibfield  {author} {\bibinfo {author} {\bibfnamefont {S.}~\bibnamefont
      {Kruk}}, \bibinfo {author} {\bibfnamefont {A.}~\bibnamefont {Slobozhanyuk}},
      \bibinfo {author} {\bibfnamefont {D.}~\bibnamefont {Denkova}}, \bibinfo
      {author} {\bibfnamefont {A.}~\bibnamefont {Poddubny}}, \bibinfo {author}
      {\bibfnamefont {I.}~\bibnamefont {Kravchenko}}, \bibinfo {author}
      {\bibfnamefont {A.}~\bibnamefont {Miroshnichenko}}, \bibinfo {author}
      {\bibfnamefont {D.}~\bibnamefont {Neshev}},\ and\ \bibinfo {author}
      {\bibfnamefont {Y.}~\bibnamefont {Kivshar}},\ }\bibfield  {title} {\bibinfo
      {title} {Edge states and topological phase transitions in chains of
      dielectric nanoparticles},\ }\href
      {https://doi.org/https://doi.org/10.1002/smll.201603190} {\bibfield
      {journal} {\bibinfo  {journal} {Small}\ }\textbf {\bibinfo {volume} {13}},\
      \bibinfo {pages} {1603190} (\bibinfo {year} {2017})},\ \Eprint
      {https://arxiv.org/abs/https://onlinelibrary.wiley.com/doi/pdf/10.1002/smll.201603190}
      {https://onlinelibrary.wiley.com/doi/pdf/10.1002/smll.201603190} \BibitemShut
      {NoStop}%
    \bibitem [{\citenamefont {Wu}\ \emph {et~al.}(2019{\natexlab{a}})\citenamefont
      {Wu}, \citenamefont {Zhang}, \citenamefont {Lee}, \citenamefont {Wang},
      \citenamefont {Yu},\ and\ \citenamefont {Fung}}]{rphwu2019}%
      \BibitemOpen
      \bibfield  {author} {\bibinfo {author} {\bibfnamefont {R.~P.~H.}\
      \bibnamefont {Wu}}, \bibinfo {author} {\bibfnamefont {Y.}~\bibnamefont
      {Zhang}}, \bibinfo {author} {\bibfnamefont {K.~F.}\ \bibnamefont {Lee}},
      \bibinfo {author} {\bibfnamefont {J.}~\bibnamefont {Wang}}, \bibinfo {author}
      {\bibfnamefont {S.~F.}\ \bibnamefont {Yu}},\ and\ \bibinfo {author}
      {\bibfnamefont {K.~H.}\ \bibnamefont {Fung}},\ }\bibfield  {title} {\bibinfo
      {title} {Dynamic long range interaction induced topological edge modes in
      dispersive gyromagnetic lattices},\ }\href
      {https://doi.org/10.1103/PhysRevB.99.214433} {\bibfield  {journal} {\bibinfo
      {journal} {Phys. Rev. B}\ }\textbf {\bibinfo {volume} {99}},\ \bibinfo
      {pages} {214433} (\bibinfo {year} {2019}{\natexlab{a}})}\BibitemShut
      {NoStop}%
    \bibitem [{\citenamefont {Asb{\'o}th}\ \emph {et~al.}(2016)\citenamefont
      {Asb{\'o}th}, \citenamefont {Oroszl{\'a}ny},\ and\ \citenamefont
      {P{\'a}lyi}}]{asboth2016}%
      \BibitemOpen
      \bibfield  {author} {\bibinfo {author} {\bibfnamefont {J.~K.}\ \bibnamefont
      {Asb{\'o}th}}, \bibinfo {author} {\bibfnamefont {L.}~\bibnamefont
      {Oroszl{\'a}ny}},\ and\ \bibinfo {author} {\bibfnamefont {A.}~\bibnamefont
      {P{\'a}lyi}},\ }\bibinfo {title} {The su-schrieffer-heeger (ssh) model},\ in\
      \href {https://doi.org/10.1007/978-3-319-25607-8_1} {\emph {\bibinfo
      {booktitle} {A Short Course on Topological Insulators: Band Structure and
      Edge States in One and Two Dimensions}}}\ (\bibinfo  {publisher} {Springer
      International Publishing},\ \bibinfo {address} {Cham},\ \bibinfo {year}
      {2016})\ pp.\ \bibinfo {pages} {1--22}\BibitemShut {NoStop}%
    \bibitem [{\citenamefont {St-Jean}\ \emph {et~al.}(2017)\citenamefont
      {St-Jean}, \citenamefont {Goblot}, \citenamefont {Galopin}, \citenamefont
      {Lema{\^\i}tre}, \citenamefont {Ozawa}, \citenamefont {Le~Gratiet},
      \citenamefont {Sagnes}, \citenamefont {Bloch},\ and\ \citenamefont
      {Amo}}]{st2017}%
      \BibitemOpen
      \bibfield  {author} {\bibinfo {author} {\bibfnamefont {P.}~\bibnamefont
      {St-Jean}}, \bibinfo {author} {\bibfnamefont {V.}~\bibnamefont {Goblot}},
      \bibinfo {author} {\bibfnamefont {E.}~\bibnamefont {Galopin}}, \bibinfo
      {author} {\bibfnamefont {A.}~\bibnamefont {Lema{\^\i}tre}}, \bibinfo {author}
      {\bibfnamefont {T.}~\bibnamefont {Ozawa}}, \bibinfo {author} {\bibfnamefont
      {L.}~\bibnamefont {Le~Gratiet}}, \bibinfo {author} {\bibfnamefont
      {I.}~\bibnamefont {Sagnes}}, \bibinfo {author} {\bibfnamefont
      {J.}~\bibnamefont {Bloch}},\ and\ \bibinfo {author} {\bibfnamefont
      {A.}~\bibnamefont {Amo}},\ }\bibfield  {title} {\bibinfo {title} {Lasing in
      topological edge states of a one-dimensional lattice},\ }\href@noop {}
      {\bibfield  {journal} {\bibinfo  {journal} {Nature Photon.}\ }\textbf
      {\bibinfo {volume} {11}},\ \bibinfo {pages} {651} (\bibinfo {year}
      {2017})}\BibitemShut {NoStop}%
    \bibitem [{\citenamefont {Zhao}\ \emph {et~al.}(2018)\citenamefont {Zhao},
      \citenamefont {Miao}, \citenamefont {Teimourpour}, \citenamefont {Malzard},
      \citenamefont {El-Ganainy}, \citenamefont {Schomerus},\ and\ \citenamefont
      {Feng}}]{zhao2018}%
      \BibitemOpen
      \bibfield  {author} {\bibinfo {author} {\bibfnamefont {H.}~\bibnamefont
      {Zhao}}, \bibinfo {author} {\bibfnamefont {P.}~\bibnamefont {Miao}}, \bibinfo
      {author} {\bibfnamefont {M.~H.}\ \bibnamefont {Teimourpour}}, \bibinfo
      {author} {\bibfnamefont {S.}~\bibnamefont {Malzard}}, \bibinfo {author}
      {\bibfnamefont {R.}~\bibnamefont {El-Ganainy}}, \bibinfo {author}
      {\bibfnamefont {H.}~\bibnamefont {Schomerus}},\ and\ \bibinfo {author}
      {\bibfnamefont {L.}~\bibnamefont {Feng}},\ }\bibfield  {title} {\bibinfo
      {title} {Topological hybrid silicon microlasers},\ }\href@noop {} {\bibfield
      {journal} {\bibinfo  {journal} {Nat. Commun.}\ }\textbf {\bibinfo {volume}
      {9}},\ \bibinfo {pages} {981} (\bibinfo {year} {2018})}\BibitemShut {NoStop}%
    \bibitem [{\citenamefont {Parto}\ \emph {et~al.}(2018)\citenamefont {Parto},
      \citenamefont {Wittek}, \citenamefont {Hodaei}, \citenamefont {Harari},
      \citenamefont {Bandres}, \citenamefont {Ren}, \citenamefont {Rechtsman},
      \citenamefont {Segev}, \citenamefont {Christodoulides},\ and\ \citenamefont
      {Khajavikhan}}]{parto2018}%
      \BibitemOpen
      \bibfield  {author} {\bibinfo {author} {\bibfnamefont {M.}~\bibnamefont
      {Parto}}, \bibinfo {author} {\bibfnamefont {S.}~\bibnamefont {Wittek}},
      \bibinfo {author} {\bibfnamefont {H.}~\bibnamefont {Hodaei}}, \bibinfo
      {author} {\bibfnamefont {G.}~\bibnamefont {Harari}}, \bibinfo {author}
      {\bibfnamefont {M.~A.}\ \bibnamefont {Bandres}}, \bibinfo {author}
      {\bibfnamefont {J.}~\bibnamefont {Ren}}, \bibinfo {author} {\bibfnamefont
      {M.~C.}\ \bibnamefont {Rechtsman}}, \bibinfo {author} {\bibfnamefont
      {M.}~\bibnamefont {Segev}}, \bibinfo {author} {\bibfnamefont {D.~N.}\
      \bibnamefont {Christodoulides}},\ and\ \bibinfo {author} {\bibfnamefont
      {M.}~\bibnamefont {Khajavikhan}},\ }\bibfield  {title} {\bibinfo {title}
      {Edge-mode lasing in 1d topological active arrays},\ }\href
      {https://doi.org/10.1103/PhysRevLett.120.113901} {\bibfield  {journal}
      {\bibinfo  {journal} {Phys. Rev. Lett.}\ }\textbf {\bibinfo {volume} {120}},\
      \bibinfo {pages} {113901} (\bibinfo {year} {2018})}\BibitemShut {NoStop}%
    \bibitem [{\citenamefont {Guo}\ \emph {et~al.}(2021)\citenamefont {Guo},
      \citenamefont {Zhang}, \citenamefont {Song}, \citenamefont {Jiang},\ and\
      \citenamefont {Chen}}]{guo2021}%
      \BibitemOpen
      \bibfield  {author} {\bibinfo {author} {\bibfnamefont {Z.}~\bibnamefont
      {Guo}}, \bibinfo {author} {\bibfnamefont {T.}~\bibnamefont {Zhang}}, \bibinfo
      {author} {\bibfnamefont {J.}~\bibnamefont {Song}}, \bibinfo {author}
      {\bibfnamefont {H.}~\bibnamefont {Jiang}},\ and\ \bibinfo {author}
      {\bibfnamefont {H.}~\bibnamefont {Chen}},\ }\bibfield  {title} {\bibinfo
      {title} {Sensitivity of topological edge states in a non-hermitian dimer
      chain},\ }\href {https://doi.org/10.1364/PRJ.413873} {\bibfield  {journal}
      {\bibinfo  {journal} {Photon. Res.}\ }\textbf {\bibinfo {volume} {9}},\
      \bibinfo {pages} {574} (\bibinfo {year} {2021})}\BibitemShut {NoStop}%
    \bibitem [{\citenamefont {Ryu}\ \emph {et~al.}(2003)\citenamefont {Ryu},
      \citenamefont {Notomi},\ and\ \citenamefont {Lee}}]{ryu2003}%
      \BibitemOpen
      \bibfield  {author} {\bibinfo {author} {\bibfnamefont {H.-Y.}\ \bibnamefont
      {Ryu}}, \bibinfo {author} {\bibfnamefont {M.}~\bibnamefont {Notomi}},\ and\
      \bibinfo {author} {\bibfnamefont {Y.-H.}\ \bibnamefont {Lee}},\ }\bibfield
      {title} {\bibinfo {title} {High-quality-factor and small-mode-volume hexapole
      modes in photonic-crystal-slab nanocavities},\ }\href
      {https://doi.org/10.1063/1.1629140} {\bibfield  {journal} {\bibinfo
      {journal} {Appl. Phys. Lett.}\ }\textbf {\bibinfo {volume} {83}},\ \bibinfo
      {pages} {4294} (\bibinfo {year} {2003})},\ \Eprint
      {https://arxiv.org/abs/https://doi.org/10.1063/1.1629140}
      {https://doi.org/10.1063/1.1629140} \BibitemShut {NoStop}%
    \bibitem [{\citenamefont {Kippenberg}\ \emph {et~al.}(2004)\citenamefont
      {Kippenberg}, \citenamefont {Spillane},\ and\ \citenamefont
      {Vahala}}]{kippenberg2004}%
      \BibitemOpen
      \bibfield  {author} {\bibinfo {author} {\bibfnamefont {T.~J.}\ \bibnamefont
      {Kippenberg}}, \bibinfo {author} {\bibfnamefont {S.~M.}\ \bibnamefont
      {Spillane}},\ and\ \bibinfo {author} {\bibfnamefont {K.~J.}\ \bibnamefont
      {Vahala}},\ }\bibfield  {title} {\bibinfo {title} {Demonstration of
      ultra-high-q small mode volume toroid microcavities on a chip},\ }\href
      {https://doi.org/10.1063/1.1833556} {\bibfield  {journal} {\bibinfo
      {journal} {Appl. Phys. Lett.}\ }\textbf {\bibinfo {volume} {85}},\ \bibinfo
      {pages} {6113} (\bibinfo {year} {2004})},\ \Eprint
      {https://arxiv.org/abs/https://doi.org/10.1063/1.1833556}
      {https://doi.org/10.1063/1.1833556} \BibitemShut {NoStop}%
    \bibitem [{\citenamefont {Xiao}\ \emph {et~al.}(2010)\citenamefont {Xiao},
      \citenamefont {Li}, \citenamefont {Jiang}, \citenamefont {Hu}, \citenamefont
      {Li},\ and\ \citenamefont {Gong}}]{xiao2010}%
      \BibitemOpen
      \bibfield  {author} {\bibinfo {author} {\bibfnamefont {Y.-F.}\ \bibnamefont
      {Xiao}}, \bibinfo {author} {\bibfnamefont {B.-B.}\ \bibnamefont {Li}},
      \bibinfo {author} {\bibfnamefont {X.}~\bibnamefont {Jiang}}, \bibinfo
      {author} {\bibfnamefont {X.}~\bibnamefont {Hu}}, \bibinfo {author}
      {\bibfnamefont {Y.}~\bibnamefont {Li}},\ and\ \bibinfo {author}
      {\bibfnamefont {Q.}~\bibnamefont {Gong}},\ }\bibfield  {title} {\bibinfo
      {title} {High quality factor, small mode volume, ring-type plasmonic
      microresonator on a silver chip},\ }\href
      {https://doi.org/10.1088/0953-4075/43/3/035402} {\bibfield  {journal}
      {\bibinfo  {journal} {J. Phys. B: At. Mol. Opt. Phys.}\ }\textbf {\bibinfo
      {volume} {43}},\ \bibinfo {pages} {035402} (\bibinfo {year}
      {2010})}\BibitemShut {NoStop}%
    \bibitem [{\citenamefont {de~Leon}\ \emph {et~al.}(2012)\citenamefont
      {de~Leon}, \citenamefont {Shields}, \citenamefont {Yu}, \citenamefont
      {Englund}, \citenamefont {Akimov}, \citenamefont {Lukin},\ and\ \citenamefont
      {Park}}]{de2012}%
      \BibitemOpen
      \bibfield  {author} {\bibinfo {author} {\bibfnamefont {N.~P.}\ \bibnamefont
      {de~Leon}}, \bibinfo {author} {\bibfnamefont {B.~J.}\ \bibnamefont
      {Shields}}, \bibinfo {author} {\bibfnamefont {C.~L.}\ \bibnamefont {Yu}},
      \bibinfo {author} {\bibfnamefont {D.~E.}\ \bibnamefont {Englund}}, \bibinfo
      {author} {\bibfnamefont {A.~V.}\ \bibnamefont {Akimov}}, \bibinfo {author}
      {\bibfnamefont {M.~D.}\ \bibnamefont {Lukin}},\ and\ \bibinfo {author}
      {\bibfnamefont {H.}~\bibnamefont {Park}},\ }\bibfield  {title} {\bibinfo
      {title} {Tailoring light-matter interaction with a nanoscale plasmon
      resonator},\ }\href {https://doi.org/10.1103/PhysRevLett.108.226803}
      {\bibfield  {journal} {\bibinfo  {journal} {Phys. Rev. Lett.}\ }\textbf
      {\bibinfo {volume} {108}},\ \bibinfo {pages} {226803} (\bibinfo {year}
      {2012})}\BibitemShut {NoStop}%
    \bibitem [{\citenamefont {Seidler}\ \emph {et~al.}(2013)\citenamefont
      {Seidler}, \citenamefont {Lister}, \citenamefont {Drechsler}, \citenamefont
      {Hofrichter},\ and\ \citenamefont {St\"{o}ferle}}]{seidler2013}%
      \BibitemOpen
      \bibfield  {author} {\bibinfo {author} {\bibfnamefont {P.}~\bibnamefont
      {Seidler}}, \bibinfo {author} {\bibfnamefont {K.}~\bibnamefont {Lister}},
      \bibinfo {author} {\bibfnamefont {U.}~\bibnamefont {Drechsler}}, \bibinfo
      {author} {\bibfnamefont {J.}~\bibnamefont {Hofrichter}},\ and\ \bibinfo
      {author} {\bibfnamefont {T.}~\bibnamefont {St\"{o}ferle}},\ }\bibfield
      {title} {\bibinfo {title} {Slotted photonic crystal nanobeam cavity with an
      ultrahigh quality factor-to-mode volume ratio},\ }\href
      {https://doi.org/10.1364/OE.21.032468} {\bibfield  {journal} {\bibinfo
      {journal} {Opt. Express}\ }\textbf {\bibinfo {volume} {21}},\ \bibinfo
      {pages} {32468} (\bibinfo {year} {2013})}\BibitemShut {NoStop}%
    \bibitem [{\citenamefont {Yang}\ \emph {et~al.}(2015)\citenamefont {Yang},
      \citenamefont {Zhang}, \citenamefont {Tian}, \citenamefont {Ji},\ and\
      \citenamefont {Quan}}]{yang2015}%
      \BibitemOpen
      \bibfield  {author} {\bibinfo {author} {\bibfnamefont {D.}~\bibnamefont
      {Yang}}, \bibinfo {author} {\bibfnamefont {P.}~\bibnamefont {Zhang}},
      \bibinfo {author} {\bibfnamefont {H.}~\bibnamefont {Tian}}, \bibinfo {author}
      {\bibfnamefont {Y.}~\bibnamefont {Ji}},\ and\ \bibinfo {author}
      {\bibfnamefont {Q.}~\bibnamefont {Quan}},\ }\bibfield  {title} {\bibinfo
      {title} {Ultrahigh- $q$ and low-mode-volume parabolic radius-modulated single
      photonic crystal slot nanobeam cavity for high-sensitivity refractive index
      sensing},\ }\href {https://doi.org/10.1109/JPHOT.2015.2476761} {\bibfield
      {journal} {\bibinfo  {journal} {IEEE Photonics Journal}\ }\textbf {\bibinfo
      {volume} {7}},\ \bibinfo {pages} {1} (\bibinfo {year} {2015})}\BibitemShut
      {NoStop}%
    \bibitem [{\citenamefont {Wang}\ \emph {et~al.}(2018)\citenamefont {Wang},
      \citenamefont {Christiansen}, \citenamefont {Yu}, \citenamefont {Mørk},\
      and\ \citenamefont {Sigmund}}]{wang2018}%
      \BibitemOpen
      \bibfield  {author} {\bibinfo {author} {\bibfnamefont {F.}~\bibnamefont
      {Wang}}, \bibinfo {author} {\bibfnamefont {R.~E.}\ \bibnamefont
      {Christiansen}}, \bibinfo {author} {\bibfnamefont {Y.}~\bibnamefont {Yu}},
      \bibinfo {author} {\bibfnamefont {J.}~\bibnamefont {Mørk}},\ and\ \bibinfo
      {author} {\bibfnamefont {O.}~\bibnamefont {Sigmund}},\ }\bibfield  {title}
      {\bibinfo {title} {Maximizing the quality factor to mode volume ratio for
      ultra-small photonic crystal cavities},\ }\href
      {https://doi.org/10.1063/1.5064468} {\bibfield  {journal} {\bibinfo
      {journal} {Appl. Phys. Lett.}\ }\textbf {\bibinfo {volume} {113}},\ \bibinfo
      {pages} {241101} (\bibinfo {year} {2018})},\ \Eprint
      {https://arxiv.org/abs/https://doi.org/10.1063/1.5064468}
      {https://doi.org/10.1063/1.5064468} \BibitemShut {NoStop}%
    \bibitem [{\citenamefont {Wu}\ \emph {et~al.}(2019{\natexlab{b}})\citenamefont
      {Wu}, \citenamefont {Wang}, \citenamefont {Chen}, \citenamefont {Chen},
      \citenamefont {Li}, \citenamefont {Tong},\ and\ \citenamefont
      {Fan}}]{xwu2019}%
      \BibitemOpen
      \bibfield  {author} {\bibinfo {author} {\bibfnamefont {X.}~\bibnamefont
      {Wu}}, \bibinfo {author} {\bibfnamefont {Y.}~\bibnamefont {Wang}}, \bibinfo
      {author} {\bibfnamefont {Q.}~\bibnamefont {Chen}}, \bibinfo {author}
      {\bibfnamefont {Y.-C.}\ \bibnamefont {Chen}}, \bibinfo {author}
      {\bibfnamefont {X.}~\bibnamefont {Li}}, \bibinfo {author} {\bibfnamefont
      {L.}~\bibnamefont {Tong}},\ and\ \bibinfo {author} {\bibfnamefont
      {X.}~\bibnamefont {Fan}},\ }\bibfield  {title} {\bibinfo {title} {High-q,
      low-mode-volume microsphere-integrated fabry\&\#x2013;perot cavity for
      optofluidic lasing applications},\ }\href
      {https://doi.org/10.1364/PRJ.7.000050} {\bibfield  {journal} {\bibinfo
      {journal} {Photon. Res.}\ }\textbf {\bibinfo {volume} {7}},\ \bibinfo {pages}
      {50} (\bibinfo {year} {2019}{\natexlab{b}})}\BibitemShut {NoStop}%
    \bibitem [{\citenamefont {Zhen}\ \emph {et~al.}(2014)\citenamefont {Zhen},
      \citenamefont {Hsu}, \citenamefont {Lu}, \citenamefont {Stone},\ and\
      \citenamefont {Solja\ifmmode \check{c}\else
      \v{c}\fi{}i\ifmmode~\acute{c}\else \'{c}\fi{}}}]{zhen2014}%
      \BibitemOpen
      \bibfield  {author} {\bibinfo {author} {\bibfnamefont {B.}~\bibnamefont
      {Zhen}}, \bibinfo {author} {\bibfnamefont {C.~W.}\ \bibnamefont {Hsu}},
      \bibinfo {author} {\bibfnamefont {L.}~\bibnamefont {Lu}}, \bibinfo {author}
      {\bibfnamefont {A.~D.}\ \bibnamefont {Stone}},\ and\ \bibinfo {author}
      {\bibfnamefont {M.}~\bibnamefont {Solja\ifmmode \check{c}\else
      \v{c}\fi{}i\ifmmode~\acute{c}\else \'{c}\fi{}}},\ }\bibfield  {title}
      {\bibinfo {title} {Topological nature of optical bound states in the
      continuum},\ }\href {https://doi.org/10.1103/PhysRevLett.113.257401}
      {\bibfield  {journal} {\bibinfo  {journal} {Phys. Rev. Lett.}\ }\textbf
      {\bibinfo {volume} {113}},\ \bibinfo {pages} {257401} (\bibinfo {year}
      {2014})}\BibitemShut {NoStop}%
    \bibitem [{\citenamefont {Hsu}\ \emph {et~al.}(2016)\citenamefont {Hsu},
      \citenamefont {Zhen}, \citenamefont {Stone}, \citenamefont {Joannopoulos},\
      and\ \citenamefont {Solja{\v{c}}i{\'c}}}]{hsu2016}%
      \BibitemOpen
      \bibfield  {author} {\bibinfo {author} {\bibfnamefont {C.~W.}\ \bibnamefont
      {Hsu}}, \bibinfo {author} {\bibfnamefont {B.}~\bibnamefont {Zhen}}, \bibinfo
      {author} {\bibfnamefont {A.~D.}\ \bibnamefont {Stone}}, \bibinfo {author}
      {\bibfnamefont {J.~D.}\ \bibnamefont {Joannopoulos}},\ and\ \bibinfo {author}
      {\bibfnamefont {M.}~\bibnamefont {Solja{\v{c}}i{\'c}}},\ }\bibfield  {title}
      {\bibinfo {title} {Bound states in the continuum},\ }\href@noop {} {\bibfield
       {journal} {\bibinfo  {journal} {Nat. Rev. Mater.}\ }\textbf {\bibinfo
      {volume} {1}},\ \bibinfo {pages} {16048} (\bibinfo {year}
      {2016})}\BibitemShut {NoStop}%
    \bibitem [{\citenamefont {Doeleman}\ \emph {et~al.}(2018)\citenamefont
      {Doeleman}, \citenamefont {Monticone}, \citenamefont {den Hollander},
      \citenamefont {Al{\`u}},\ and\ \citenamefont {Koenderink}}]{doeleman2018}%
      \BibitemOpen
      \bibfield  {author} {\bibinfo {author} {\bibfnamefont {H.~M.}\ \bibnamefont
      {Doeleman}}, \bibinfo {author} {\bibfnamefont {F.}~\bibnamefont {Monticone}},
      \bibinfo {author} {\bibfnamefont {W.}~\bibnamefont {den Hollander}}, \bibinfo
      {author} {\bibfnamefont {A.}~\bibnamefont {Al{\`u}}},\ and\ \bibinfo {author}
      {\bibfnamefont {A.~F.}\ \bibnamefont {Koenderink}},\ }\bibfield  {title}
      {\bibinfo {title} {Experimental observation of a polarization vortex at an
      optical bound state in the continuum},\ }\href@noop {} {\bibfield  {journal}
      {\bibinfo  {journal} {Nature Photon.}\ }\textbf {\bibinfo {volume} {12}},\
      \bibinfo {pages} {397} (\bibinfo {year} {2018})}\BibitemShut {NoStop}%
    \bibitem [{\citenamefont {Pankin}\ \emph {et~al.}(2020)\citenamefont {Pankin},
      \citenamefont {Wu}, \citenamefont {Yang}, \citenamefont {Chen}, \citenamefont
      {Timofeev},\ and\ \citenamefont {Sadreev}}]{pankin2020}%
      \BibitemOpen
      \bibfield  {author} {\bibinfo {author} {\bibfnamefont {P.}~\bibnamefont
      {Pankin}}, \bibinfo {author} {\bibfnamefont {B.-R.}\ \bibnamefont {Wu}},
      \bibinfo {author} {\bibfnamefont {J.-H.}\ \bibnamefont {Yang}}, \bibinfo
      {author} {\bibfnamefont {K.-P.}\ \bibnamefont {Chen}}, \bibinfo {author}
      {\bibfnamefont {I.}~\bibnamefont {Timofeev}},\ and\ \bibinfo {author}
      {\bibfnamefont {A.}~\bibnamefont {Sadreev}},\ }\bibfield  {title} {\bibinfo
      {title} {One-dimensional photonic bound states in the continuum},\
      }\href@noop {} {\bibfield  {journal} {\bibinfo  {journal} {Commun. Phys.}\
      }\textbf {\bibinfo {volume} {3}},\ \bibinfo {pages} {91} (\bibinfo {year}
      {2020})}\BibitemShut {NoStop}%
    \bibitem [{\citenamefont {Azzam}\ and\ \citenamefont
      {Kildishev}(2021)}]{azzam2021}%
      \BibitemOpen
      \bibfield  {author} {\bibinfo {author} {\bibfnamefont {S.~I.}\ \bibnamefont
      {Azzam}}\ and\ \bibinfo {author} {\bibfnamefont {A.~V.}\ \bibnamefont
      {Kildishev}},\ }\bibfield  {title} {\bibinfo {title} {Photonic bound states
      in the continuum: From basics to applications},\ }\href
      {https://doi.org/https://doi.org/10.1002/adom.202001469} {\bibfield
      {journal} {\bibinfo  {journal} {Adv. Optical Mater.}\ }\textbf {\bibinfo
      {volume} {9}},\ \bibinfo {pages} {2001469} (\bibinfo {year} {2021})},\
      \Eprint
      {https://arxiv.org/abs/https://onlinelibrary.wiley.com/doi/pdf/10.1002/adom.202001469}
      {https://onlinelibrary.wiley.com/doi/pdf/10.1002/adom.202001469} \BibitemShut
      {NoStop}%
    \bibitem [{\citenamefont {Kuttge}\ \emph {et~al.}(2010)\citenamefont {Kuttge},
      \citenamefont {García~de Abajo},\ and\ \citenamefont {Polman}}]{kuttge2010}%
      \BibitemOpen
      \bibfield  {author} {\bibinfo {author} {\bibfnamefont {M.}~\bibnamefont
      {Kuttge}}, \bibinfo {author} {\bibfnamefont {F.~J.}\ \bibnamefont {García~de
      Abajo}},\ and\ \bibinfo {author} {\bibfnamefont {A.}~\bibnamefont {Polman}},\
      }\bibfield  {title} {\bibinfo {title} {Ultrasmall mode volume plasmonic
      nanodisk resonators},\ }\href {https://doi.org/10.1021/nl902546r} {\bibfield
      {journal} {\bibinfo  {journal} {Nano Lett.}\ }\textbf {\bibinfo {volume}
      {10}},\ \bibinfo {pages} {1537} (\bibinfo {year} {2010})},\ \bibinfo {note}
      {pMID: 19813755},\ \Eprint
      {https://arxiv.org/abs/https://doi.org/10.1021/nl902546r}
      {https://doi.org/10.1021/nl902546r} \BibitemShut {NoStop}%
    \bibitem [{\citenamefont {Huang}\ \emph {et~al.}(2016)\citenamefont {Huang},
      \citenamefont {Ming}, \citenamefont {Lin}, \citenamefont {Ling},
      \citenamefont {Ruan}, \citenamefont {Palacios}, \citenamefont {Wang},
      \citenamefont {Dresselhaus},\ and\ \citenamefont {Kong}}]{huang2016}%
      \BibitemOpen
      \bibfield  {author} {\bibinfo {author} {\bibfnamefont {S.}~\bibnamefont
      {Huang}}, \bibinfo {author} {\bibfnamefont {T.}~\bibnamefont {Ming}},
      \bibinfo {author} {\bibfnamefont {Y.}~\bibnamefont {Lin}}, \bibinfo {author}
      {\bibfnamefont {X.}~\bibnamefont {Ling}}, \bibinfo {author} {\bibfnamefont
      {Q.}~\bibnamefont {Ruan}}, \bibinfo {author} {\bibfnamefont {T.}~\bibnamefont
      {Palacios}}, \bibinfo {author} {\bibfnamefont {J.}~\bibnamefont {Wang}},
      \bibinfo {author} {\bibfnamefont {M.}~\bibnamefont {Dresselhaus}},\ and\
      \bibinfo {author} {\bibfnamefont {J.}~\bibnamefont {Kong}},\ }\bibfield
      {title} {\bibinfo {title} {Ultrasmall mode volumes in plasmonic cavities of
      nanoparticle-on-mirror structures},\ }\href
      {https://doi.org/https://doi.org/10.1002/smll.201601318} {\bibfield
      {journal} {\bibinfo  {journal} {Small}\ }\textbf {\bibinfo {volume} {12}},\
      \bibinfo {pages} {5190} (\bibinfo {year} {2016})},\ \Eprint
      {https://arxiv.org/abs/https://onlinelibrary.wiley.com/doi/pdf/10.1002/smll.201601318}
      {https://onlinelibrary.wiley.com/doi/pdf/10.1002/smll.201601318} \BibitemShut
      {NoStop}%
    \bibitem [{\citenamefont {Hugall}\ \emph {et~al.}(2018)\citenamefont {Hugall},
      \citenamefont {Singh},\ and\ \citenamefont {van Hulst}}]{hugall2018}%
      \BibitemOpen
      \bibfield  {author} {\bibinfo {author} {\bibfnamefont {J.~T.}\ \bibnamefont
      {Hugall}}, \bibinfo {author} {\bibfnamefont {A.}~\bibnamefont {Singh}},\ and\
      \bibinfo {author} {\bibfnamefont {N.~F.}\ \bibnamefont {van Hulst}},\
      }\bibfield  {title} {\bibinfo {title} {Plasmonic cavity coupling},\ }\href
      {https://doi.org/10.1021/acsphotonics.7b01139} {\bibfield  {journal}
      {\bibinfo  {journal} {ACS Photonics}\ }\textbf {\bibinfo {volume} {5}},\
      \bibinfo {pages} {43} (\bibinfo {year} {2018})},\ \Eprint
      {https://arxiv.org/abs/https://doi.org/10.1021/acsphotonics.7b01139}
      {https://doi.org/10.1021/acsphotonics.7b01139} \BibitemShut {NoStop}%
    \bibitem [{\citenamefont {Epstein}\ \emph {et~al.}(2020)\citenamefont
      {Epstein}, \citenamefont {Alcaraz}, \citenamefont {Huang}, \citenamefont
      {Pusapati}, \citenamefont {Hugonin}, \citenamefont {Kumar}, \citenamefont
      {Deputy}, \citenamefont {Khodkov}, \citenamefont {Rappoport}, \citenamefont
      {Hong}, \citenamefont {Peres}, \citenamefont {Kong}, \citenamefont {Smith},\
      and\ \citenamefont {Koppens}}]{epstein2020}%
      \BibitemOpen
      \bibfield  {author} {\bibinfo {author} {\bibfnamefont {I.}~\bibnamefont
      {Epstein}}, \bibinfo {author} {\bibfnamefont {D.}~\bibnamefont {Alcaraz}},
      \bibinfo {author} {\bibfnamefont {Z.}~\bibnamefont {Huang}}, \bibinfo
      {author} {\bibfnamefont {V.-V.}\ \bibnamefont {Pusapati}}, \bibinfo {author}
      {\bibfnamefont {J.-P.}\ \bibnamefont {Hugonin}}, \bibinfo {author}
      {\bibfnamefont {A.}~\bibnamefont {Kumar}}, \bibinfo {author} {\bibfnamefont
      {X.~M.}\ \bibnamefont {Deputy}}, \bibinfo {author} {\bibfnamefont
      {T.}~\bibnamefont {Khodkov}}, \bibinfo {author} {\bibfnamefont {T.~G.}\
      \bibnamefont {Rappoport}}, \bibinfo {author} {\bibfnamefont {J.-Y.}\
      \bibnamefont {Hong}}, \bibinfo {author} {\bibfnamefont {N.~M.~R.}\
      \bibnamefont {Peres}}, \bibinfo {author} {\bibfnamefont {J.}~\bibnamefont
      {Kong}}, \bibinfo {author} {\bibfnamefont {D.~R.}\ \bibnamefont {Smith}},\
      and\ \bibinfo {author} {\bibfnamefont {F.~H.~L.}\ \bibnamefont {Koppens}},\
      }\bibfield  {title} {\bibinfo {title} {Far-field excitation of single
      graphene plasmon cavities with ultracompressed mode volumes},\ }\href
      {https://doi.org/10.1126/science.abb1570} {\bibfield  {journal} {\bibinfo
      {journal} {Science}\ }\textbf {\bibinfo {volume} {368}},\ \bibinfo {pages}
      {1219} (\bibinfo {year} {2020})},\ \Eprint
      {https://arxiv.org/abs/https://www.science.org/doi/pdf/10.1126/science.abb1570}
      {https://www.science.org/doi/pdf/10.1126/science.abb1570} \BibitemShut
      {NoStop}%
    \bibitem [{\citenamefont {Han}\ \emph {et~al.}(2009)\citenamefont {Han},
      \citenamefont {Lai}, \citenamefont {Fung}, \citenamefont {Zhang},\ and\
      \citenamefont {Chan}}]{han2009}%
      \BibitemOpen
      \bibfield  {author} {\bibinfo {author} {\bibfnamefont {D.}~\bibnamefont
      {Han}}, \bibinfo {author} {\bibfnamefont {Y.}~\bibnamefont {Lai}}, \bibinfo
      {author} {\bibfnamefont {K.~H.}\ \bibnamefont {Fung}}, \bibinfo {author}
      {\bibfnamefont {Z.-Q.}\ \bibnamefont {Zhang}},\ and\ \bibinfo {author}
      {\bibfnamefont {C.~T.}\ \bibnamefont {Chan}},\ }\bibfield  {title} {\bibinfo
      {title} {Negative group velocity from quadrupole resonance of plasmonic
      spheres},\ }\href {https://doi.org/10.1103/PhysRevB.79.195444} {\bibfield
      {journal} {\bibinfo  {journal} {Phys. Rev. B}\ }\textbf {\bibinfo {volume}
      {79}},\ \bibinfo {pages} {195444} (\bibinfo {year} {2009})}\BibitemShut
      {NoStop}%
    \bibitem [{\citenamefont {Al\`u}\ and\ \citenamefont
      {Engheta}(2009)}]{alu2009}%
      \BibitemOpen
      \bibfield  {author} {\bibinfo {author} {\bibfnamefont {A.}~\bibnamefont
      {Al\`u}}\ and\ \bibinfo {author} {\bibfnamefont {N.}~\bibnamefont
      {Engheta}},\ }\bibfield  {title} {\bibinfo {title} {Guided propagation along
      quadrupolar chains of plasmonic nanoparticles},\ }\href
      {https://doi.org/10.1103/PhysRevB.79.235412} {\bibfield  {journal} {\bibinfo
      {journal} {Phys. Rev. B}\ }\textbf {\bibinfo {volume} {79}},\ \bibinfo
      {pages} {235412} (\bibinfo {year} {2009})}\BibitemShut {NoStop}%
    \bibitem [{\citenamefont {Evlyukhin}\ \emph {et~al.}(2012)\citenamefont
      {Evlyukhin}, \citenamefont {Reinhardt}, \citenamefont {Zywietz},\ and\
      \citenamefont {Chichkov}}]{evlyukhin2012}%
      \BibitemOpen
      \bibfield  {author} {\bibinfo {author} {\bibfnamefont {A.~B.}\ \bibnamefont
      {Evlyukhin}}, \bibinfo {author} {\bibfnamefont {C.}~\bibnamefont
      {Reinhardt}}, \bibinfo {author} {\bibfnamefont {U.}~\bibnamefont {Zywietz}},\
      and\ \bibinfo {author} {\bibfnamefont {B.~N.}\ \bibnamefont {Chichkov}},\
      }\bibfield  {title} {\bibinfo {title} {Collective resonances in metal
      nanoparticle arrays with dipole-quadrupole interactions},\ }\href
      {https://doi.org/10.1103/PhysRevB.85.245411} {\bibfield  {journal} {\bibinfo
      {journal} {Phys. Rev. B}\ }\textbf {\bibinfo {volume} {85}},\ \bibinfo
      {pages} {245411} (\bibinfo {year} {2012})}\BibitemShut {NoStop}%
    \bibitem [{\citenamefont {Swiecicki}\ and\ \citenamefont
      {Sipe}(2017)}]{swiecicki2017}%
      \BibitemOpen
      \bibfield  {author} {\bibinfo {author} {\bibfnamefont {S.~D.}\ \bibnamefont
      {Swiecicki}}\ and\ \bibinfo {author} {\bibfnamefont {J.~E.}\ \bibnamefont
      {Sipe}},\ }\bibfield  {title} {\bibinfo {title} {Surface-lattice resonances
      in two-dimensional arrays of spheres: Multipolar interactions and a mode
      analysis},\ }\href {https://doi.org/10.1103/PhysRevB.95.195406} {\bibfield
      {journal} {\bibinfo  {journal} {Phys. Rev. B}\ }\textbf {\bibinfo {volume}
      {95}},\ \bibinfo {pages} {195406} (\bibinfo {year} {2017})}\BibitemShut
      {NoStop}%
    \bibitem [{\citenamefont {Babicheva}\ and\ \citenamefont
      {Evlyukhin}(2018)}]{babicheva2018}%
      \BibitemOpen
      \bibfield  {author} {\bibinfo {author} {\bibfnamefont {V.~E.}\ \bibnamefont
      {Babicheva}}\ and\ \bibinfo {author} {\bibfnamefont {A.~B.}\ \bibnamefont
      {Evlyukhin}},\ }\bibfield  {title} {\bibinfo {title} {Metasurfaces with
      electric quadrupole and magnetic dipole resonant coupling},\ }\href
      {https://doi.org/10.1021/acsphotonics.7b01520} {\bibfield  {journal}
      {\bibinfo  {journal} {ACS Photonics}\ }\textbf {\bibinfo {volume} {5}},\
      \bibinfo {pages} {2022} (\bibinfo {year} {2018})},\ \Eprint
      {https://arxiv.org/abs/https://doi.org/10.1021/acsphotonics.7b01520}
      {https://doi.org/10.1021/acsphotonics.7b01520} \BibitemShut {NoStop}%
    \bibitem [{\citenamefont {Babicheva}\ and\ \citenamefont
      {Evlyukhin}(2019)}]{babicheva2019}%
      \BibitemOpen
      \bibfield  {author} {\bibinfo {author} {\bibfnamefont {V.~E.}\ \bibnamefont
      {Babicheva}}\ and\ \bibinfo {author} {\bibfnamefont {A.~B.}\ \bibnamefont
      {Evlyukhin}},\ }\bibfield  {title} {\bibinfo {title} {Analytical model of
      resonant electromagnetic dipole-quadrupole coupling in nanoparticle arrays},\
      }\href {https://doi.org/10.1103/PhysRevB.99.195444} {\bibfield  {journal}
      {\bibinfo  {journal} {Phys. Rev. B}\ }\textbf {\bibinfo {volume} {99}},\
      \bibinfo {pages} {195444} (\bibinfo {year} {2019})}\BibitemShut {NoStop}%
    \bibitem [{\citenamefont {Novotny}\ and\ \citenamefont
      {Hecht}(2012)}]{novotny2012}%
      \BibitemOpen
      \bibfield  {author} {\bibinfo {author} {\bibfnamefont {L.}~\bibnamefont
      {Novotny}}\ and\ \bibinfo {author} {\bibfnamefont {B.}~\bibnamefont
      {Hecht}},\ }\bibinfo {title} {Theoretical foundations},\ in\ \href
      {https://doi.org/10.1017/CBO9780511794193.004} {\emph {\bibinfo {booktitle}
      {Principles of Nano-Optics}}}\ (\bibinfo  {publisher} {Cambridge University
      Press},\ \bibinfo {year} {2012})\ p.\ \bibinfo {pages} {12–44},\ \bibinfo
      {edition} {2nd}\ ed.\BibitemShut {Stop}%
    \bibitem [{\citenamefont {Gon\c{c}alves}\ \emph {et~al.}(2014)\citenamefont
      {Gon\c{c}alves}, \citenamefont {Melikyan}, \citenamefont {Minassian},
      \citenamefont {Makaryan},\ and\ \citenamefont {Marti}}]{goncalves2014}%
      \BibitemOpen
      \bibfield  {author} {\bibinfo {author} {\bibfnamefont {M.~R.}\ \bibnamefont
      {Gon\c{c}alves}}, \bibinfo {author} {\bibfnamefont {A.}~\bibnamefont
      {Melikyan}}, \bibinfo {author} {\bibfnamefont {H.}~\bibnamefont {Minassian}},
      \bibinfo {author} {\bibfnamefont {T.}~\bibnamefont {Makaryan}},\ and\
      \bibinfo {author} {\bibfnamefont {O.}~\bibnamefont {Marti}},\ }\bibfield
      {title} {\bibinfo {title} {Strong dipole-quadrupole coupling and fano
      resonance in h-like metallic nanostructures},\ }\href
      {https://doi.org/10.1364/OE.22.024516} {\bibfield  {journal} {\bibinfo
      {journal} {Opt. Express}\ }\textbf {\bibinfo {volume} {22}},\ \bibinfo
      {pages} {24516} (\bibinfo {year} {2014})}\BibitemShut {NoStop}%
    \bibitem [{\citenamefont {Guo}\ \emph {et~al.}(2018)\citenamefont {Guo},
      \citenamefont {Zhang}, \citenamefont {Qian},\ and\ \citenamefont
      {Fung}}]{guo2018}%
      \BibitemOpen
      \bibfield  {author} {\bibinfo {author} {\bibfnamefont {K.}~\bibnamefont
      {Guo}}, \bibinfo {author} {\bibfnamefont {Y.-L.}\ \bibnamefont {Zhang}},
      \bibinfo {author} {\bibfnamefont {C.}~\bibnamefont {Qian}},\ and\ \bibinfo
      {author} {\bibfnamefont {K.-H.}\ \bibnamefont {Fung}},\ }\bibfield  {title}
      {\bibinfo {title} {Electric dipole-quadrupole hybridization induced
      enhancement of second-harmonic generation in t-shaped plasmonic
      heterodimers},\ }\href {https://doi.org/10.1364/OE.26.011984} {\bibfield
      {journal} {\bibinfo  {journal} {Opt. Express}\ }\textbf {\bibinfo {volume}
      {26}},\ \bibinfo {pages} {11984} (\bibinfo {year} {2018})}\BibitemShut
      {NoStop}%
    \bibitem [{\citenamefont {Pang}\ \emph {et~al.}(2021)\citenamefont {Pang},
      \citenamefont {Huang}, \citenamefont {Zhou}, \citenamefont {Mao},
      \citenamefont {Deng},\ and\ \citenamefont {Lan}}]{pang2021}%
      \BibitemOpen
      \bibfield  {author} {\bibinfo {author} {\bibfnamefont {H.}~\bibnamefont
      {Pang}}, \bibinfo {author} {\bibfnamefont {H.}~\bibnamefont {Huang}},
      \bibinfo {author} {\bibfnamefont {L.}~\bibnamefont {Zhou}}, \bibinfo {author}
      {\bibfnamefont {Y.}~\bibnamefont {Mao}}, \bibinfo {author} {\bibfnamefont
      {F.}~\bibnamefont {Deng}},\ and\ \bibinfo {author} {\bibfnamefont
      {S.}~\bibnamefont {Lan}},\ }\bibfield  {title} {\bibinfo {title} {Strong
      dipole-quadrupole-exciton coupling realized in a gold nanorod dimer placed on
      a two-dimensional material},\ }\bibfield  {journal} {\bibinfo  {journal}
      {Nanomaterials}\ }\textbf {\bibinfo {volume} {11}},\ \href
      {https://doi.org/10.3390/nano11061619} {10.3390/nano11061619} (\bibinfo
      {year} {2021})\BibitemShut {NoStop}%
    \bibitem [{\citenamefont {Bohren}\ and\ \citenamefont
      {Huffman}(1998)}]{bohren1998}%
      \BibitemOpen
      \bibfield  {author} {\bibinfo {author} {\bibfnamefont {C.~F.}\ \bibnamefont
      {Bohren}}\ and\ \bibinfo {author} {\bibfnamefont {D.~R.}\ \bibnamefont
      {Huffman}},\ }\bibinfo {title} {Particles small compared with the
      wavelength},\ in\ \href
      {https://doi.org/https://doi.org/10.1002/9783527618156.ch5} {\emph {\bibinfo
      {booktitle} {Absorption and Scattering of Light by Small Particles}}}\
      (\bibinfo  {publisher} {John Wiley \& Sons, Ltd},\ \bibinfo {year} {1998})\
      Chap.~\bibinfo {chapter} {5}, pp.\ \bibinfo {pages} {130--157},\ \Eprint
      {https://arxiv.org/abs/https://onlinelibrary.wiley.com/doi/pdf/10.1002/9783527618156.ch5}
      {https://onlinelibrary.wiley.com/doi/pdf/10.1002/9783527618156.ch5}
      \BibitemShut {NoStop}%
    \bibitem [{\citenamefont {Weber}\ and\ \citenamefont {Ford}(2004)}]{weber2004}%
      \BibitemOpen
      \bibfield  {author} {\bibinfo {author} {\bibfnamefont {W.~H.}\ \bibnamefont
      {Weber}}\ and\ \bibinfo {author} {\bibfnamefont {G.~W.}\ \bibnamefont
      {Ford}},\ }\bibfield  {title} {\bibinfo {title} {Propagation of optical
      excitations by dipolar interactions in metal nanoparticle chains},\ }\href
      {https://doi.org/10.1103/PhysRevB.70.125429} {\bibfield  {journal} {\bibinfo
      {journal} {Phys. Rev. B}\ }\textbf {\bibinfo {volume} {70}},\ \bibinfo
      {pages} {125429} (\bibinfo {year} {2004})}\BibitemShut {NoStop}%
    \bibitem [{\citenamefont {Bergman}\ and\ \citenamefont
      {Stroud}(1980)}]{bergman1980}%
      \BibitemOpen
      \bibfield  {author} {\bibinfo {author} {\bibfnamefont {D.~J.}\ \bibnamefont
      {Bergman}}\ and\ \bibinfo {author} {\bibfnamefont {D.}~\bibnamefont
      {Stroud}},\ }\bibfield  {title} {\bibinfo {title} {Theory of resonances in
      the electromagnetic scattering by macroscopic bodies},\ }\href
      {https://doi.org/10.1103/PhysRevB.22.3527} {\bibfield  {journal} {\bibinfo
      {journal} {Phys. Rev. B}\ }\textbf {\bibinfo {volume} {22}},\ \bibinfo
      {pages} {3527} (\bibinfo {year} {1980})}\BibitemShut {NoStop}%
    \bibitem [{\citenamefont {Markel}(1995)}]{markel1995}%
      \BibitemOpen
      \bibfield  {author} {\bibinfo {author} {\bibfnamefont {V.~A.}\ \bibnamefont
      {Markel}},\ }\bibfield  {title} {\bibinfo {title} {Antisymmetrical optical
      states},\ }\href {https://doi.org/10.1364/JOSAB.12.001783} {\bibfield
      {journal} {\bibinfo  {journal} {J. Opt. Soc. Am. B}\ }\textbf {\bibinfo
      {volume} {12}},\ \bibinfo {pages} {1783} (\bibinfo {year}
      {1995})}\BibitemShut {NoStop}%
    \bibitem [{\citenamefont {Fung}\ and\ \citenamefont {Chan}(2007)}]{fung2007}%
      \BibitemOpen
      \bibfield  {author} {\bibinfo {author} {\bibfnamefont {K.~H.}\ \bibnamefont
      {Fung}}\ and\ \bibinfo {author} {\bibfnamefont {C.~T.}\ \bibnamefont
      {Chan}},\ }\bibfield  {title} {\bibinfo {title} {Plasmonic modes in periodic
      metal nanoparticle chains: a direct dynamic eigenmode analysis},\ }\href
      {https://doi.org/10.1364/OL.32.000973} {\bibfield  {journal} {\bibinfo
      {journal} {Opt. Lett.}\ }\textbf {\bibinfo {volume} {32}},\ \bibinfo {pages}
      {973} (\bibinfo {year} {2007})}\BibitemShut {NoStop}%
    \bibitem [{\citenamefont {Fung}\ and\ \citenamefont {Chan}(2008)}]{fung2008}%
      \BibitemOpen
      \bibfield  {author} {\bibinfo {author} {\bibfnamefont {K.~H.}\ \bibnamefont
      {Fung}}\ and\ \bibinfo {author} {\bibfnamefont {C.~T.}\ \bibnamefont
      {Chan}},\ }\bibfield  {title} {\bibinfo {title} {Analytical study of the
      plasmonic modes of a metal nanoparticle circular array},\ }\href
      {https://doi.org/10.1103/PhysRevB.77.205423} {\bibfield  {journal} {\bibinfo
      {journal} {Phys. Rev. B}\ }\textbf {\bibinfo {volume} {77}},\ \bibinfo
      {pages} {205423} (\bibinfo {year} {2008})}\BibitemShut {NoStop}%
    \bibitem [{\citenamefont {Zhang}\ \emph {et~al.}(2020)\citenamefont {Zhang},
      \citenamefont {Wu}, \citenamefont {Shi},\ and\ \citenamefont
      {Fung}}]{zhang2020}%
      \BibitemOpen
      \bibfield  {author} {\bibinfo {author} {\bibfnamefont {Y.}~\bibnamefont
      {Zhang}}, \bibinfo {author} {\bibfnamefont {R.~P.~H.}\ \bibnamefont {Wu}},
      \bibinfo {author} {\bibfnamefont {L.}~\bibnamefont {Shi}},\ and\ \bibinfo
      {author} {\bibfnamefont {K.~H.}\ \bibnamefont {Fung}},\ }\bibfield  {title}
      {\bibinfo {title} {Second-order topological photonic modes in dipolar
      arrays},\ }\href {https://doi.org/10.1021/acsphotonics.0c00160} {\bibfield
      {journal} {\bibinfo  {journal} {ACS Photonics}\ }\textbf {\bibinfo {volume}
      {7}},\ \bibinfo {pages} {2002} (\bibinfo {year} {2020})},\ \Eprint
      {https://arxiv.org/abs/https://doi.org/10.1021/acsphotonics.0c00160}
      {https://doi.org/10.1021/acsphotonics.0c00160} \BibitemShut {NoStop}%
    \bibitem [{\citenamefont {Zak}(1989)}]{zak1989}%
      \BibitemOpen
      \bibfield  {author} {\bibinfo {author} {\bibfnamefont {J.}~\bibnamefont
      {Zak}},\ }\bibfield  {title} {\bibinfo {title} {Berry's phase for energy
      bands in solids},\ }\href {https://doi.org/10.1103/PhysRevLett.62.2747}
      {\bibfield  {journal} {\bibinfo  {journal} {Phys. Rev. Lett.}\ }\textbf
      {\bibinfo {volume} {62}},\ \bibinfo {pages} {2747} (\bibinfo {year}
      {1989})}\BibitemShut {NoStop}%
    \bibitem [{\citenamefont {Obana}\ \emph {et~al.}(2019)\citenamefont {Obana},
      \citenamefont {Liu},\ and\ \citenamefont {Wakabayashi}}]{obana2019}%
      \BibitemOpen
      \bibfield  {author} {\bibinfo {author} {\bibfnamefont {D.}~\bibnamefont
      {Obana}}, \bibinfo {author} {\bibfnamefont {F.}~\bibnamefont {Liu}},\ and\
      \bibinfo {author} {\bibfnamefont {K.}~\bibnamefont {Wakabayashi}},\
      }\bibfield  {title} {\bibinfo {title} {Topological edge states in the
      su-schrieffer-heeger model},\ }\href
      {https://doi.org/10.1103/PhysRevB.100.075437} {\bibfield  {journal} {\bibinfo
       {journal} {Phys. Rev. B}\ }\textbf {\bibinfo {volume} {100}},\ \bibinfo
      {pages} {075437} (\bibinfo {year} {2019})}\BibitemShut {NoStop}%
    \end{thebibliography}
%
    
\end{document}